\documentclass[pdflatex,sn-mathphys]{sn-jnl}

\jyear{2022}%

\theoremstyle{thmstyleone}%
\theoremstyle{thmstyletwo}%

\theoremstyle{thmstylethree}%

\begin{document}

\title[Dressed black holes in the new tensor-vector-scalar theory]{Dressed black holes in the new tensor-vector-scalar theory}


\author*{\fnm{Reginald Christian} \sur{Bernardo}}\email{rbernardo@gate.sinica.edu.tw}
\equalcont{These authors contributed equally to this work.}

\author{\fnm{Che-Yu} \sur{Chen}}\email{b97202056@gmail.com}
\equalcont{These authors contributed equally to this work.}


\affil{\orgdiv{Institute of Physics}, \orgname{Academia Sinica}, \orgaddress{\city{Taipei}, \postcode{11529}, \country{Taiwan}}}




\abstract{As incarnations of gravity in its prime, black holes are arguably the best target for us to demystify gravity. Keeping in mind the prominent role black holes play in gravitational wave astronomy, it becomes a must for a theory to possess black hole solutions with only measurable departures from their general relativity counterparts. In this paper, we present black holes in a tensor-vector-scalar representation of relativistic modified Newtonian dynamics. We find that the theory allows Schwarzschild and nearly-Schwarzschild black holes as solutions, while the nontrivial scalar and vector fields generally diverge at the event horizon. Whether this is a physical pathology or not poses a challenge for these solutions, and by extension, the model. However, even if it is, this pathology could be overcome when the black hole hair vanishes.}




\maketitle


\section{Introduction}
\label{sec:intro}

Black holes embody gravity. They stand on their own stability, supporting themselves through the nonlinear nature of gravity, and thus hold the key to understand the fate of the Universe. Gravitational wave astronomy \cite{LIGOScientific:2016aoc, LIGOScientific:2020kqk} and black hole imaging \cite{EventHorizonTelescope:2019dse, EventHorizonTelescope:2019ggy} support further black holes as ubiquitous astrophysical objects. This opens up a golden opportunity to study the dynamical nature of black holes in a rich physical environment, provided there are also theories and predictions to compare observations with.

General relativity (GR) comes with a pleasant surprise, that black holes have no hair, i.e., they are completely described by mass, charge, and spin. This is the content of the celebrated no-hair theorems \cite{Israel:1967wq, Israel:1967za, Carter:1971zc}. Outside of GR, an expectation is therefore for black holes to have hair, or be surrounded by nontrivial fields. It comes as an amazing counterstatement that in alternative gravity, black holes can also often be geometrically described by their GR solution, except with or without hair \cite{Bekenstein:1995un, Hertog:2006rr, Hui:2012qt, Sotiriou:2015pka}. This important result implies that black holes can be recognized not just as probes of strong gravity but also laboratories for studying the existence of additional fields, such as dark matter and dark energy which stand out due to their cosmological relevance \cite{Barack:2018yly, Barausse:2020rsu}. Looking out for black holes in alternative gravity is tied with the longstanding scientific goal of understanding these dark fields.

Modified Newtonian dynamics (MoND) \cite{Milgrom:1983ca, Milgrom:1983pn, Milgrom:1983zz} tackled the dark matter paradigm by invoking a threshold acceleration below which the gravitational force on a particle drops as an inverse distance, \textit{not} inverse-square. This explained the tail end dynamics and the distribution of dark matter surrounding galaxies. However, despite its simplicity and phenomenological prowess, the lack of a covariant theory of MoND made it challenging for the model to be tested in the relativistic cosmological limit and was frequently ground for it to be considered less seriously. Admirable attempts to construct a relativistic MoND were nonetheless pursued which eventually ushered progress \cite{Bekenstein:2004ne, Sanders:2005vd, Skordis:2005xk, Zlosnik:2006sb, Babichev:2011kq, Famaey:2011kh, Zlosnik:2017xpr}. A tensor-vector-scalar (TeVeS) theory (Eq. \eqref{eq:rmond}) was recently motivated to be a consistent relativistic modified Newtonian dynamics (RMoND) \cite{PhysRevD.100.104013, Skordis:2020eui}. Cosmological and weak field limits were shown to be compatible with observations in this model \cite{Skordis:2020eui, Skordis:2021vuk}; however, the strong gravity regime remains to be given further attention. We fill in this gap in this TeVeS theory by studying its black hole solutions.

Gravitational wave constraints on the post-Newtonian parameters continue to support GR, and by extension, its black holes as an incredible  portrayal of astrophysical compact objects in binary mergers \cite{LIGOScientific:2021sio}. This implies that a Schwarzschild black hole may be taken as a first approximation of an astrophysical black hole in the nonrotating limit. Nontrivial fields (dark or visible) on a Schwarzschild background therefore reflect whether a theory can be of use phenomenologically in the strong gravity regime. Keeping this in mind, we look at the hairy Schwarzschild and nearly-Schwarzschild black holes in the TeVeS theory. 

This paper proceeds as follows. We introduce the action and the covariant field equations of TeVeS (Section \ref{sec:new_mond}). We present its field equations in the strong gravity regime (Section \ref{sec:strong_gravity}). We work in isotropic coordinates where the post-Newtonian parameters can be read off easily from just the metric. We obtain the spherically symmetric solutions in the model, which represent black hole spacetimes and non-black hole objects (Sections \ref{sec:time_pointed_vector}, \ref{sec:on_vanishing_charge}, and \ref{sec:time_space_vector}). We pay attention to hairy Schwarzschild black holes and in addition present a designer algorithm for solutions of the system. We discuss the implication of our results, particularly on the fields that stand on the Schwarzschild background (Section \ref{sec:outlook}).

We work with the mostly-plus metric signature $(-, +, +, +)$ and in geometrized units $\left( c = 8 \pi \tilde{G} = 1 \right)$ where $c$ is the speed of light in vacuum and $\tilde{G}$ is Newton's gravitational constant. We acknowledge the ``xAct'' and its derivatives ``xPert'' and ``xCoba'' packages for our calculations. Our calculations are transparently communicated in a Mathematica notebook which can be downloaded from \href{https://github.com/reggiebernardo/notebooks}{GitHub} \cite{reggie_bernardo_4810864}.

\section{The tensor-vector-scalar theory}
\label{sec:new_mond}

We consider the TeVeS theory whose \textit{gravitational} action is given by \cite{Skordis:2020eui, Kashfi:2022dyb}
\begin{equation}
\label{eq:rmond}
\begin{split}
S_g = \int d^4 x \dfrac{\sqrt{-g}}{16 \pi \tilde{G}} \bigg[ \mathcal{R} & - \dfrac{K_B}{2} F^{ab} F_{ab} + 2 \left( 2 - K_B \right) J^a \nabla_a \phi \\
& - \left( 2 - K_B \right) Y - F \left( Y, Q \right) - \lambda \left( A^a A_a + 1 \right) \bigg]\,,
\end{split}
\end{equation}
where $g_{ab}$ is the metric, $g$ is the determinant of $g_{ab}$, $\mathcal{R}$ is the Ricci scalar, $\phi$ is the scalar field, $A^b$ is the vector field with the kinetic term $F_{ab} = 2 \nabla_{[a} A_{b]} = \nabla_{a} A_b - \nabla_b A_a$, $J_a = A^b \nabla_b A_a$, $Y = q^{ab} \nabla_a \phi \nabla_b \phi$ where $q_{ab} = g_{ab} + A_a A_b$, and $Q = A^a \nabla_a \phi$. In addition, $K_B$ is a coupling constant, $\lambda$ is a Lagrange multiplier enforcing the vector field to be timelike, and $F$ is an arbitrary function of its two arguments. We refer to $F$ as the TeVeS potential. In terms of the scalar field, it is also worth noting that the theory is invariant under the shift symmetry $\phi \rightarrow \phi + \phi_0$ and comes with gravitational waves travelling at the speed of light. Matter fields $\Psi$ can be considered by adding a matter action $S_m[g, \Psi]$ such that $\delta S_m[g, \Psi] / \delta g^{ab} = - 2 T^{(\text{Mat})}_{ab} / \sqrt{-g}$ where $T^{(\text{Mat})}_{ab}$ is the matter stress-energy tensor.

The field equations can be obtained by varying the action \eqref{eq:rmond} with respect to the fields $(g_{ab}, \phi, A^a)$ and multiplier $\lambda$. By performing the metric variation, we obtain the modified Einstein equation
\begin{equation}
\label{eq:einstein_eq}
G_{ab} - T_{ab}^{(\phi,A)} = 8 \pi \tilde{G} T^{(\text{Mat})}_{ab}\,,
\end{equation}
where $T_{ab}^{(\phi, A)}$ captures the MoND degrees of freedom and is given by
\begin{equation}
\label{eq:set_mond}
\begin{split}
    T_{ab}^{(\phi, A)} = \ & \dfrac{g_{ab}}{2} \bigg[ - \dfrac{K_B}{2} F^{ij} F_{ij} + (2 - K_B) \left(2 J^i \nabla_i \phi - {Y} \right) - {F} \bigg] \\
    & \phantom{gg} + K_B F_{a}^{\ j}F_{bj} + {F}_{Q} A_{(a} \nabla_{b)} \phi + \lambda A_a A_b \\
    & \phantom{gg} - 2 (2 - K_B) \bigg[ A_{(a} \left( \nabla_{b)} A^j \right) \left( \nabla_j \phi \right) + A^{j} \left( \nabla_j A_{(a} \right) \left( \nabla_{b)} \phi \right) \bigg] \\
    & \phantom{gg} -2 (2 - K_B) \bigg[ - \nabla_{(i} \left( g_{j)(a} A_{b)} A^i \nabla^j \phi \right) + \dfrac{1}{2} \nabla_j \left( A^j A_{(a} \nabla_{b)} \phi \right) \bigg] \\
    & \phantom{gg} + (2 - K_B + {F}_{Y}) \bigg[ \left( \nabla_a \phi \right) \left( \nabla_b \phi \right) + 2 A_{(a} \left( \nabla_{b)} \phi \right) A^j \nabla_j \phi \bigg] \,.
\end{split}
\end{equation}
Subscripts ${Y}$ or ${Q}$ on the potential ${F}$ denote differentiation with respect to ${Y}$ or ${Q}$, e.g., ${F}_{Y} = \partial {F} / \partial {Y}$, ${F}_{Q} = \partial {F} / \partial {Q}$, and ${F}_{{Y} {Q}} = \partial^2 {F} / \partial {Y} \partial {Q}$. On the other hand, the variation with respect to the scalar field gives
\begin{equation}
\label{eq:sfe}
\nabla_a S^a = 0 \,,
\end{equation}
where
\begin{equation}
\label{eq:shift_current}
S^a = 2 \left( 2 - K_B \right) \left( J^a - q^{ab} \nabla_b \phi \right) - 2 {F}_{{Y}} q^{ab} \nabla_b \phi - {F}_{Q} A^a \,,
\end{equation}
while the variation with respect to the vector field leads to
\begin{equation}
\label{eq:vfe}
\begin{split}
0 = 2 K_B \nabla_a F^{ab} & - 2 {F}_{{Y}} A^a \nabla_a \phi \nabla^b \phi - {F}_{{Q}} \nabla^b \phi - 2 \lambda A^b \\
+ 2 ( 2 & - K_B ) \bigg[ \nabla^b A^c \nabla_c \phi - \nabla_c \left( A^c \nabla^b \phi \right) - A^c \nabla_c \phi \nabla^b \phi \bigg] \,.
\end{split}
\end{equation}
In addition to these dynamical field equations, varying the multiplier leads to the constraint equation
\begin{equation}
\label{eq:lambda_constraint}
A^b A_b = -1
\end{equation}
for the vector field. This simply implies that the dynamics will be on the hypersurface where the vector field is unit timelike.

We spend this paper on the static and spherically symmetric solutions of the TeVeS system (Eqs. \eqref{eq:einstein_eq}, \eqref{eq:sfe}, \eqref{eq:vfe}, and \eqref{eq:lambda_constraint}) in vacuum ($T_{ab}^{(\text{Mat})} = 0$). In GR, this configuration ansatz naturally gives rise to the Schwarzschild solution and it represents the strong gravity regime of the theory. We now turn our attention to this regime in the TeVeS theory.

\section{The strong gravity regime}
\label{sec:strong_gravity}

We present the strong gravity regime of TeVeS by considering its static and spherically symmetric vacuum solutions. We setup the field equations in isotropic coordinates which will be solved in the following sections.

We consider the metric in isotropic coordinates \cite{Giannios:2005es}
\begin{equation}
\label{eq:iso_coords}
ds^2 = - e^{\nu(r)} dt^2 + e^{\zeta(r)} \left( dr^2 + r^2 d\Omega^2 \right)\,,
\end{equation}
where $d\Omega^2 = d\theta^2 + \sin \theta^2 d \varphi^2$ is the line element on the unit two-sphere, and $\nu(r)$, $\zeta(r)$ are the metric functions. In addition, we write down the scalar field as
\begin{equation}
\label{eq:phiq0}
\phi = \Phi(r)\,,
\end{equation}
and the vector field to be
\begin{equation}
A_b = \left( {U}(r), {V}(r), 0, 0 \right) \,.
\end{equation}
The scalar and vector fields respect the staticity and spherical symmetry of the spacetime.

The $tt$- and $rr$-components of the modified Einstein equation (Eq. \eqref{eq:einstein_eq}) are given by
\begin{equation}
\label{eq:tt}
\begin{split}
E_{tt}\left[ \nu, \zeta, \Phi, {U}, {V}, \lambda \right] = 0\,,
\end{split}
\end{equation}
and
\begin{equation}
\label{eq:rr}
\begin{split}
E_{rr}\left[ \nu, \zeta, \Phi, {U}, {V}, \lambda \right] = 0 \,,
\end{split}
\end{equation}
respectively, where the explicit functional expressions of $E_{ab}$ are given by Eqs. \eqref{eq:Ett} and \eqref{eq:Err} in Appendix \ref{sec:explicit_field_equations}. A prime denotes differentiation with respect to $r$. It is also useful to consider the $tr$- and $\theta\theta$-components
\begin{equation}
\label{eq:tr}
\begin{split}
E_{tr}\left[ \nu, \zeta, \Phi, {U}, {V}, \lambda \right] = 0\,,
\end{split}
\end{equation}
and
\begin{equation}
\label{eq:thetatheta}
\begin{split}
E_{\theta\theta}\left[ \nu, \zeta, \Phi, {U}, {V}, \lambda \right] = 0 \,,
\end{split}
\end{equation}
respectively, where $E_{tr}$ and $E_{\theta\theta}$ are given by Eqs. \eqref{eq:Etr} and \eqref{eq:Ethetatheta}. On the other hand, we write down the $r$-component of the shift current (Eq. \eqref{eq:shift_current}) as
\begin{equation}
\label{eq:Sr}
\begin{split}
S^r = & -2 e^{-2 \zeta } {V}^2 \Phi ' {F}_{{Y}}-e^{-\zeta } {V} {F}_{{Q}}-2 e^{-\zeta } \Phi ' {F}_{{Y}} \\
& \phantom{gg} +2 e^{-\zeta } K_B \Phi ' -4 e^{-\zeta } \Phi ' \\
& \phantom{gg} - (K_B - 2) {U}^2 e^{-\zeta -\nu } \nu ' + (K_B - 2) e^{-2 \zeta } {V}^2 \zeta ' \\
& \phantom{gg} +2 e^{-2 \zeta } (K_B - 2) {V}^2 \Phi '-2 e^{-2 \zeta } (K_B - 2) {V} {V}' \,,
\end{split}
\end{equation}
through which the scalar field equation (Eq. \eqref{eq:sfe}) can be simply written as
\begin{equation}
\label{eq:sfe_iso}
\partial_r \left( e^{ (3 \zeta + \nu)/2 } r^2 S^r \right) = 0 \,.
\end{equation}
Lastly, the $t$- and $r$-components of the vector field equation (Eq. \eqref{eq:vfe}) are given by
\begin{equation}
\label{eq:vfe_t}
\begin{split}
0 = & 2 K_B r {U}''+K_B {U}' \left(r \zeta '-r \nu '+4\right) +2 r {U} \left((K_B-2) \nu ' \Phi '-e^{\zeta } \lambda \right)\,,
\end{split}
\end{equation}
and
\begin{equation}
\label{eq:vfe_r}
\begin{split}
0 = & -e^{\zeta } r \left(\Phi ' {F}_{Q}+2 \lambda  {V}\right) + {V} \bigg( -2 r \left(\Phi '\right)^2 {F}_{Y} \\
& \phantom{gg} + (K_B-2) \left(\Phi ' \left(r \left(\zeta '+\nu '+2 \Phi '\right)+4\right)+2 r \Phi ''\right) \bigg)\,,
\end{split}
\end{equation}
respectively. The scalar-vector couplings become
\begin{equation}
\label{eq:Y_iso}
{Y} = e^{-2 \zeta } \left(e^{\zeta }+{V}^2\right) \left(\Phi '\right)^2\,,
\end{equation}
and
\begin{equation}
\label{eq:Q_iso}
{Q} = e^{-\zeta } {V} \Phi '\,,
\end{equation}
and the timelike vector constraint becomes
\begin{equation}
\label{eq:timelike_v_iso}
e^{-\nu } {U}^2-e^{-\zeta } {V}^2-1=0 \,.
\end{equation}
We look for the solutions of the system of equations (Eqs. \eqref{eq:tt}, \eqref{eq:rr}, \eqref{eq:tr}, \eqref{eq:thetatheta}, \eqref{eq:sfe_iso}, \eqref{eq:vfe_t}, and \eqref{eq:vfe_r}) for the tensor, vector, and scalar degrees of freedom. It should be noted that with spherical symmetry, $G_{\theta\theta} = G_{\varphi\varphi}/\sin^2\theta$, and one of the Bianchi identities, namely, $\nabla_a G^{ab} = 0$, only two of the components of the modified Einstein equation need to be considered. But we shall keep the $\theta\theta$ component as an additional resource for simplifying the system of equations.

{Before closing this section, we would like to briefly mention the possibility of considering the more general ansatz $\phi = q t + \Phi(r)$, which is permitted by the shift symmetry within the scalar field sector. The main challenges with $q \neq 0$ are that, apart from the field equations' becoming too unwieldy, there are more equations that had to be dealt with\footnote{{The interested reader may check them and possibly extend the work using our publicly available \href{https://github.com/reggiebernardo/notebooks/tree/main/supp_ntbks_arxiv.2202.08460}{codes}} \cite{reggie_bernardo_4810864}.}. Granted, in beyond Horndeski theory, the additional $tr$ component of the modified Einstein equation is proportional to the scalar field equation, but whether this is also true in TeVeS is still uncertain. Therefore, we opted to focus on the $q = 0$ branch to make first insights to the strong gravity regime of the theory.}

\section{{Vector field aligned along the time direction}}
\label{sec:time_pointed_vector}

We first consider the vector field to be pointing directly along the time direction, i.e., ${V} = 0$ and ${U} \neq 0$ for all $r$. In this case, the field equations can be simplified significantly; particularly, the $tr$-component of the modified Einstein equation (Eq. \eqref{eq:tr}) and the $r$-component of the vector field equation (Eq. \eqref{eq:vfe_r}) reduce to $\Phi' {F}_{Q} = 0$. In the branch $\Phi' = 0$, the scalar-vector couplings merely vanish and so the system becomes identical to that of GR with a cosmological constant related to ${F}(0, 0)$, provided that $K_B = 0$. This limit remains to be phenomenologically interesting when considering perturbations of the solution. The branch $\Phi' = 0$ with $K_B \neq 0$ also remains nontrivial due to the vector field retaining its kinetic term. We explore this in \ref{sec:unit_v_sol}. However, in this paper, we are instead interested in the nontrivial branch with $\Phi' \neq 0$ and therefore set ${F}_{Q} = 0$ in what follows.

We are left with the $tt$-, $rr$-, $\theta\theta$-components of the modified Einstein equation, the scalar field equation (Eq. \eqref{eq:sfe}), the time component of the vector field equation (Eq. \eqref{eq:vfe_t}), and the unit-timelike vector constraint (Eq. \eqref{eq:timelike_v_iso}). Notably, the system also still depends on the potential ${F}$ that functionally captures how the metric would respond to the presence of nontrivial hair.

\subsection{Analytical solutions}
\label{subsec:analytical}

In this subsection, we present two exact analytical solutions: a non-black hole spacetime solution corresponding to $K_B = 2$, and a hairy Schwarzschild black hole for ${F} \sim Y$. In the next subsection, we obtain necessary conditions for the metric functions and present an algorithm for designing solutions in TeVeS.

\subsubsection*{A non-black hole solution}

It can be shown that
\begin{equation}
\label{eq:k2_metric}
    ds^2 = - e^{\nu(r)} dt^2 + \dfrac{e^{-\nu(r)}}{r^4} \left( dr^2 + r^2 d\Omega^2 \right)\,,
\end{equation}
together with a constant scalar field $\Phi$ and a time-pointing vector field is an exact solution to the system with ${F}(0, 0) = 0$ and $K_B = 2$. This covers analytical functions of the potential ${F}$ and retains the kinetic term of the vector field and the nontrivial couplings. The function $\nu(r)$ is unconstrained and the multiplier is determined by the field equations as
\begin{equation}
    \lambda = \dfrac{1}{4} K_B r^4 e^{\nu (r)} \left(\nu '(r)^2-2 \nu ''(r)\right) \,.
\end{equation}
This solution does not represent a black hole. To understand this, one could transform to Schwarzschild coordinates using $\chi^2 = e^{-\nu(r)}/r^2$ where the line element becomes
\begin{equation}
    ds^2 = -e^{\nu(r)} dt^2 + \dfrac{d\chi^2}{ \left( \left(r \nu'(r)/2\right) + 1 \right)^2 } + \chi^2 d\Omega^2 \,.
\end{equation}
This form makes it clear that at infinity, $\chi \rightarrow \infty$, this spacetime can be asymptotically flat, if $e^{\nu} \sim 1$. To pass as a black hole solution, however, both functions $e^{\nu}$ and $\left(r \nu'/2\right) + 1$ must vanish simultaneously and at equal rates at the event horizon $\chi = \chi_H$ (or in isotropic coordinates $r = r_H$). We cannot find a function that satisfies this as $\nu$ must necessarily diverge in order for $e^{\nu}$ to vanish. Nonetheless, this spacetime remains nontrivial and illustrative as a solution to the theory.

We emphasize that what we refer to as ``non-black holes'' are horizonless objects. They can be compact, as they are spherical yet supported in the scalar-vector-tensor gravitational vacuum. In addition, the solutions could be asymptotically flat and free from naked singularities. For example, the Ricci scalar (Eq. \eqref{eq:ricci}) corresponding to the metric \eqref{eq:k2_metric} is given by $\mathcal{R} \sim \nu'(r)^2 - 2 \nu''(r)$. An exact $\mathcal{R} = 0$ solution can be singled out with $e^{\nu(r)} = 1 / \left( r + 2 L \right)^2$ where $L > 0$ is an arbitrary length scale. In this case, the Kretschmann scalar (Eq. \eqref{eq:kretschmann}) becomes $\mathcal{K} = 8 r^6 \left( r^2 + 24L^2 \right)/\left( r + 2L \right)^8$. Obviously, the curvature is bounded everywhere, $\mathcal{K} < \infty$ for any $r \geq 0$, which shows that the spacetime is free from curvature singularities. Therefore, these objects can still be very interesting from both theoretical and astrophysical points of view. However, since we will be mainly focusing on black hole spacetimes in this paper, the discussion regarding these horizonless non-black hole solutions will be postponed to elsewhere.

\subsubsection*{A hairy Schwarzschild black hole}

The Schwarzschild black hole is given by
\begin{equation}
\label{eq:schw_nu}
e^{\nu(r)} = \left( \dfrac{1 - R/(4r)}{1 + R/(4r)} \right)^2\,,
\end{equation}
and
\begin{equation}
\label{eq:schw_zeta}
e^{\zeta(r)} = \left( 1 + R/(4r) \right)^4\,,
\end{equation}
with $R$ a constant quantifying the black hole event horizon at $r = R/4$. It is well known that this is an exact solution in GR which can be inspected by retaining only the Einstein-Hilbert term in the action. On the other hand, this comes as a solution to the system of equations together with a scalar hair given by
\begin{equation}
\label{eq:phi_hairy_schw}
    \Phi' \left(r\right) = \dfrac{K_B R}{2 (K_B - 2) \left( r^2 - (R/4)^2 \right)}\,,
\end{equation}
and the TeVeS potential
\begin{equation}
\label{eq:Fpot_hairy_schw}
    {F({Y}, {Q})} = 2 \left( K_B - 2 \right) {Y} / K_B \,.
\end{equation}
The multiplier that comes with this is given by
\begin{equation}
    \lambda = \dfrac{1024 K_B r^4 R^2}{\left(r - (R/4) \right)^2 (4 r+R)^6} \,.
\end{equation}
This is a hairy Schwarzschild black hole with both nontrivial scalar and vector fields. Note that if $K_B = 0$, or rather when both scalar and vector couplings vanish, we recover GR.

\subsection{Necessary conditions}
\label{subsec:model_independent}

We derive necessary conditions\footnote{{Our usage of `necessary condition' concurs with the physics language, which is an equation derived from some `parent' equations. For example, the wave equation is a necessary condition of the Maxwell equations, meaning that the fields propagate in media. Solutions of Maxwell equations always satisfy the wave equation, but not the other way around, that is, not all propagating waves are solutions of the Maxwell equations, unless the electric and magnetic fields are transverse with each other and the wave direction. We take advantage of the necessary condition nonetheless by substituting its solution back to the parent equations.}} for the independent variables $(\nu, \zeta, \Phi)$. This is in preparation for the designer approach which we use to obtain exact solutions (Sections \ref{subsec:nearly_schw} and \ref{subsec:designer_recipe}).

For a time-pointing vector field, i.e., ${V} = 0$, we can eliminate the vector field by using Eq. \eqref{eq:timelike_v_iso} to obtain the constraint
\begin{equation}
\label{eq:vector_vto}
{U}(r) = e^{\nu(r)/2} \,. 
\end{equation}
By substituting this into the field equations, we obtain a coupled system for the variables $\nu$, $\zeta$, and $\Phi$, but also the Lagrange multiplier $\lambda$. The multiplier can be eliminated by using the remaining nontrivial time component of the vector field equation. This leads to
\begin{equation}
\label{eq:multiplier_vto}
\lambda = \dfrac{e^{-\zeta}}{4 r} \left(\nu' \left(K_B r \zeta'+4 \left(K_B+(K_B-2) r \Phi'\right)\right)+2 K_B r \nu''\right) \,.
\end{equation}
By substituting this into the modified Einstein and scalar field equations, we obtain a system explicitly for just the metric functions $\left( \nu , \zeta \right)$ and the scalar field $\Phi$. To be concrete, the $tt$-, $rr$-, and $\theta\theta$-components of the modified Einstein equation become
\begin{equation}
\label{eq:tt_vto}
\begin{split}
0 = & -4 e^{\zeta } r {F}-8 K_B \nu '+16 K_B \Phi '-8 r \zeta '' \\
& \ \ -2 \zeta ' \left(K_B r \nu '-2 (K_B-2) r \Phi '+8\right) \\
& \ \ -2 r \left(\zeta '\right)^2 -4 K_B r \nu ''-K_B r \left(\nu '\right)^2 -32 \Phi ' \\
& \ \ +4 K_B r \left(\Phi '\right)^2 -16 r \Phi ''-8 r \left(\Phi '\right)^2 +8 K_B r \Phi ''\,,
\end{split}
\end{equation}
\begin{equation}
\label{eq:rr_vto}
\begin{split}
0 = & -4 \left(\Phi '\right)^2 {F}_{Y}+2 e^{\zeta } {F}+\left(\zeta '\right)^2 \\
& \phantom{gg} -2 K_B \nu ' \Phi '+\frac{1}{2} K_B \left(\nu '\right)^2+2 K_B \left(\Phi '\right)^2 \\
& \phantom{gg} +4 \nu ' \Phi '+\frac{2 \zeta ' \left(r \nu '+2\right)}{r}+\frac{4 \nu '}{r}-4 \left(\Phi '\right)^2 \,,
\end{split}
\end{equation}
and
\begin{equation}
\label{eq:thetatheta_vto}
\begin{split}
0 = \ & 4 e^{\zeta } r {F}+4 \zeta '+4 \nu '+4 r \zeta '' \\
& \phantom{gg} +4 K_B r \nu ' \Phi ' -K_B r \left(\nu '\right)^2-4 K_B r \left(\Phi '\right)^2 \\
& \phantom{gg} +4 r \nu ''-8 r \nu ' \Phi '+2 r \left(\nu '\right)^2+8 r \left(\Phi '\right)^2 \,,
\end{split}
\end{equation}
respectively, while the scalar field equation (Eq. \eqref{eq:sfe_iso}) can be integrated to obtain
\begin{equation}
\label{eq:sfe_vto}
2 r \Phi ' \left({F}_{Y}-K_B+2\right)+(K_B-2) r \nu '= - \frac{{J}}{r} e^{-(\nu + \zeta)/2} \,.
\end{equation}
In Eq. \eqref{eq:sfe_vto}, the quantity ${J}$ is an integration constant that we refer to as a shift charge. When ${J} = 0$, then the scalar field equation reduces to $S^r = 0$; in general, Eq. \eqref{eq:sfe_vto} corresponds to $\exp\left( (3 \zeta + \nu)/2 \right) r^2 S^r = \text{constant}$ with the constant proportional to ${J}$.

A notable observation in Eqs. \eqref{eq:tt_vto}, \eqref{eq:rr_vto}, \eqref{eq:thetatheta_vto}, and \eqref{eq:sfe_vto} is that they depend on the arbitrary potential only explicitly through ${F}$ and ${F}_{Y}$. Therefore, we can utilize two of the above equations to eliminate the potential ${F}$ dependence in the remaining equations. We particularly isolate ${F}$ in the $\theta\theta$-component and ${F}_{Y}$ in the scalar field equation. Substituting these into the $tt$- and $rr$-components leads to the following ${F}$-independent equations
\begin{equation}
\label{eq:tt_vto_mi}
\begin{split}
0 = \ & 2 ( r ( \zeta ''+(K_B-1) \nu '' \\
& \phantom{gggggi} -2 (K_B-2) \Phi '' ) -4 (K_B-2) \Phi' ) \\
& \phantom{gg} +\zeta ' \left(K_B r \nu '-2 (K_B-2) r \Phi '+6\right) +r \left(\zeta '\right)^2 \\
& \phantom{gg} +\nu ' \left(4 K_B-2 (K_B-2) r \Phi '-2\right)+(K_B-1) r \left(\nu '\right)^2\,,
\end{split}
\end{equation}
and
\begin{equation}
\label{eq:rr_vto_mi}
\begin{split}
2 r \left(\zeta ''+\nu ''\right) = \ & 2 \zeta ' \left(r \nu '+1\right)+r \left(\zeta '\right)^2 \\
& \phantom{gg} +\nu ' \left(2-2 (K_B-2) r \Phi '\right) \\
& \phantom{gg} +(K_B-1) r \left(\nu '\right)^2 +\frac{2 {J} e^{-(\nu + \zeta)/2} \Phi '}{r} \,,
\end{split}
\end{equation}
respectively. Eqs. \eqref{eq:tt_vto_mi} and \eqref{eq:rr_vto_mi} can be interpreted as necessary conditions for a solution $\left( \nu, \zeta, \Phi \right)$ of the system of equations. A step further can be obtained by recognizing that Eq. \eqref{eq:rr_vto_mi} is an algebraic equation for $\Phi'$. However, the resulting equation is not presentable in the general case with a nonvanishing shift charge, i.e., ${J} \neq 0$. Thus, we focus on the vanishing shift charge limit for the illustrative purposes of this work. By solving Eq. \eqref{eq:rr_vto_mi} for $\Phi'$ and substituting into Eq. \eqref{eq:tt_vto_mi}, we obtain
\begin{equation}
\label{eq:ttrr_vto_mi_J0}
\begin{split}
0 = \ & r^2 \zeta ^{\prime 3} \nu '-2 \bigg( -2 r^2 \nu '' \left(\zeta ''+\nu ''\right) +\nu^{\prime 2} \left(r^2 \nu ''-2\right) \\
& \phantom{gggggggggiii} +2 r \nu ' \left(\zeta ''+2 \nu ''+r \left(\zeta'''+\nu''' \right)\right) \bigg) \\
& \phantom{gg} +\zeta ' \bigg( 2 \nu ' \left(r^2 \zeta ''-r^2 \nu ''+2\right) +r^2 \nu^{\prime 3}-4 r \nu ''+6 r \nu^{\prime 2} \bigg) \\
&  \phantom{gg} +2 r \zeta^{\prime 2} \left(3 \nu '-r \nu ''+r \nu^{\prime 2}\right) \,.
\end{split}
\end{equation}
We refer to Eq. \eqref{eq:ttrr_vto_mi_J0} as a necessary condition explicitly for the metric functions $\nu$ and $\zeta$ only. This is for example satisfied by the Schwarzschild solution (Eqs. \eqref{eq:schw_nu} and \eqref{eq:schw_zeta}). We shall use Eq. \eqref{eq:ttrr_vto_mi_J0} as a starting point of a solution-generating algorithm and consider deviations from the Schwarzschild metric in order to find hairy black holes in this TeVeS representation of RMoND. 

We highlight two things about Eq. \eqref{eq:ttrr_vto_mi_J0}. First, the appearance of a metric-only necessary condition such as Eq. \eqref{eq:ttrr_vto_mi_J0} that is satisfied by the GR solution is known in other alternative gravity models. In particular, similar situations have appeared in Horndeski theory \cite{Bernardo:2019mmx} and modified teleparallel gravity \cite{Bahamonde:2021gfp} when studying the strong gravity regime. Second, Eq. \eqref{eq:ttrr_vto_mi_J0} is a second-order nonlinear differential equation in the first-derivatives $\nu'$ and $\zeta'$. Obviously, this cannot be solved (as we have not specified the free potential ${F}$) and we are left with no more independent equations to spare. Getting around this therefore requires specifying either of the metric functions, and determining the other one by solving a second-order nonlinear differential equation. In this way, the potential ${F}$ (corresponding to a choice of either $\nu$ or $\zeta$) can be determined by tracing the steps back to the ${F}$-dependent field equations. We demonstrate this designer approach in the remainder of this section.

\subsection{Nearly-Schwarzschild black holes in TeVeS}
\label{subsec:nearly_schw}

We construct nearly-Schwarzschild solutions to detail a solution-generating method inspired from a designer algorithm in Horndeski theory \cite{Bernardo:2019mmx}. As alluded to previously, one either provides ${F}$ (closing the system of equations) and solves for $(\nu, \zeta, \Phi)$, or alternatively specifies one of the metric functions ($\nu$ or $\zeta$) and traces the necessary conditions backward to determine $(\nu, \Phi, {F})$ (or $(\zeta, \Phi, {F})$). The former is the canonical way of treating modified gravity with arbitrary potentials. We draw the designing approach from the latter.

To initiate the procedure, we opt to specify $\zeta$ as given by the Schwarzschild metric function (Eq. \eqref{eq:schw_zeta}). This then turns Eq. \eqref{eq:ttrr_vto_mi_J0} into a differential equation for the metric function $\nu$: ${D}[\nu(r)] = 0$. However, rather than specifying $\nu$ by its Schwarzschild functional form (Eq. \eqref{eq:schw_nu}), we solve the ${D}[\nu(r)] = 0$ by imposing Schwarzschild asymptotics at $r \gg R$. Expanding Eq. \eqref{eq:schw_nu} in powers of $R/r$ leads to
\begin{equation}
\label{eq:schw_nu_asympt}
    \nu(r \gg R) = -\dfrac{R}{r} - \dfrac{R^3}{48 r^3} + O \left( (R/r)^{5} \right) \,.
\end{equation}
We consider two cases: one which is using only the leading order term (first term) in Eq. \eqref{eq:schw_nu_asympt}, and the other one including the next-to-leading order (first two terms). The physical motivation for this choice is mostly clear as we are working in isotropic coordinates. Since $\zeta$, or rather the spatial component of the metric, is fixed to exactly its GR value, the post-Newtonian parameter $\gamma$ is exactly unity. On the other hand, the post-Newtonian parameter $\beta$ modulating the redshift function $e^\nu = 1 - (2M/r) + ( 2\beta M^2/r^2) + \cdots$ can be completely determined by the leading order (LO) term in the asymptotics \eqref{eq:schw_nu_asympt} and is exactly unity as well. The next-to-leading order (NLO) in \eqref{eq:schw_nu_asympt} would contribute to higher orders post-Newtonian parameters.

\begin{figure}[h!]
\center
\includegraphics[width = 0.8 \textwidth]{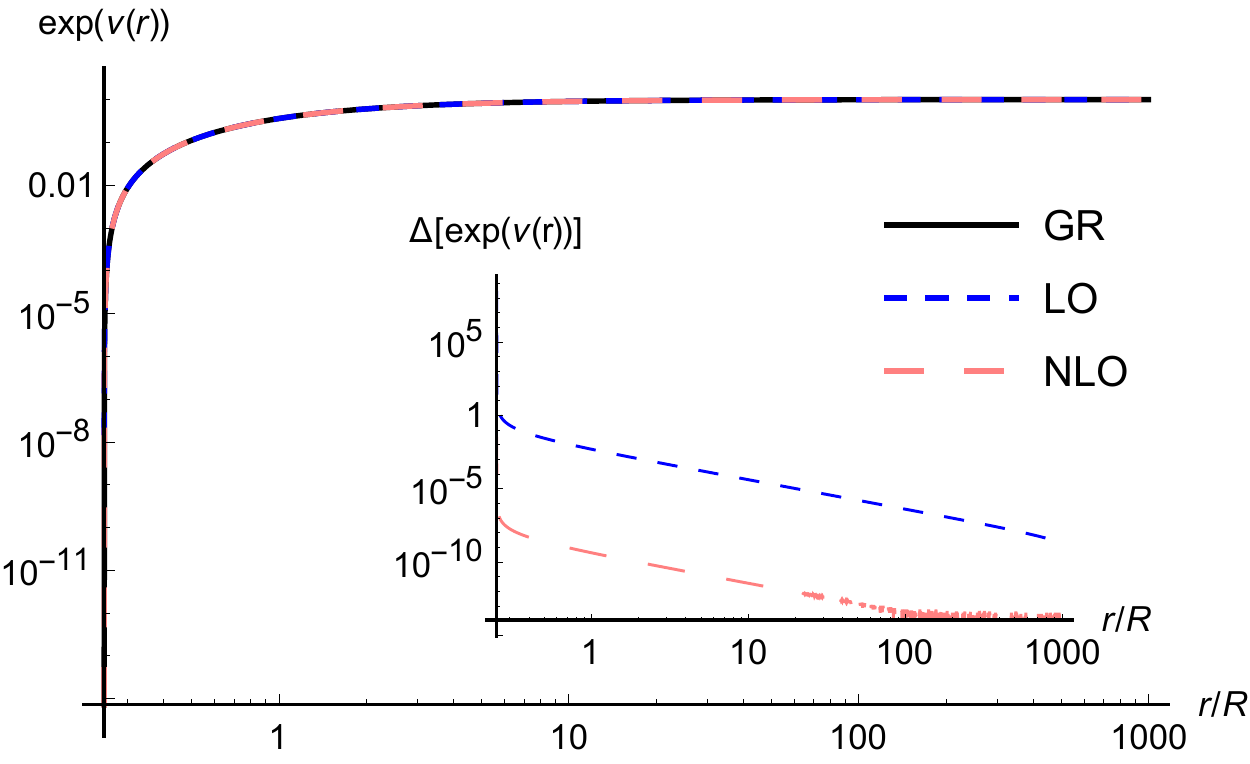}
\caption{The redshift function $\exp( \nu(r) )$ solved through Eq. \eqref{eq:ttrr_vto_mi_J0} together with the Schwarzschild asymptotic condition Eq. \eqref{eq:schw_nu_asympt} imposed at $r/R = 10^3$. LO (NLO) stands for (next-to-)leading order which means the first (two) term(s) in Eq. \eqref{eq:schw_nu_asympt} is (are) considered. The inset shows the percent deviation of the LO and NLO solutions compared to the Schwarzschild solution.}
\label{fig:metric_near_schw}
\end{figure}

Starting the numerical integration at $r/R = 10^3$, and then integrating toward the Schwarzschild event horizon $r \sim R/4$, we obtain the solutions in Figure \ref{fig:metric_near_schw}.

In both cases, the solutions can be seen to be visually indistinguishable from the Schwarzschild solution (Eq. \eqref{eq:schw_nu}). Only careful inspection shows minute differences particularly as one goes away from the starting point of the integration. This is supported by the inset of Figure \ref{fig:metric_near_schw} showing the percent deviation of the LO and NLO solutions compared with the Schwarzschild redshift function. Understandably, the deviation of the NLO solution is smaller compared to the LO one. Regardless, both cases share the same shape as far as the deviation goes. The percent deviation increases away from the integration initial point and becomes even particularly considerable as the solution approaches the Schwarzschild event horizon $r \sim R/4$. This is notably of order unity in the LO case.

We must highlight that the event horizon $r_H$ for the LO and NLO cases turn out to be below the Schwarzschild case $r_\text{Schw} = R/4$. The change of the event horizon position is $\delta r_H/R = (r_H - r_\text{Schw})/R = -6 \times 10^{-5}$ for the LO case while $\delta r_H/R = -6 \times 10^{-12}$ for the NLO. The negative sign means these new horizons are below the Schwarzschild horizon. This also implies that the curves of the LO and NLO are above the GR case in Figure \ref{fig:metric_near_schw} as their horizons are below that of the corresponding Schwarzschild black hole. In order, at $r \sim R/4$, $e^{\nu(r)}_\text{LO} > e^{\nu(r)}_\text{NLO} > e^{\nu(r)}_\text{GR} = 0$ for the solutions in the vicinity of the horizon. We were careful in examining these minute differences, keeping our numerical precision to 20 digits throughout.

\begin{figure}[h!]
\center
\includegraphics[width = 0.8 \textwidth]{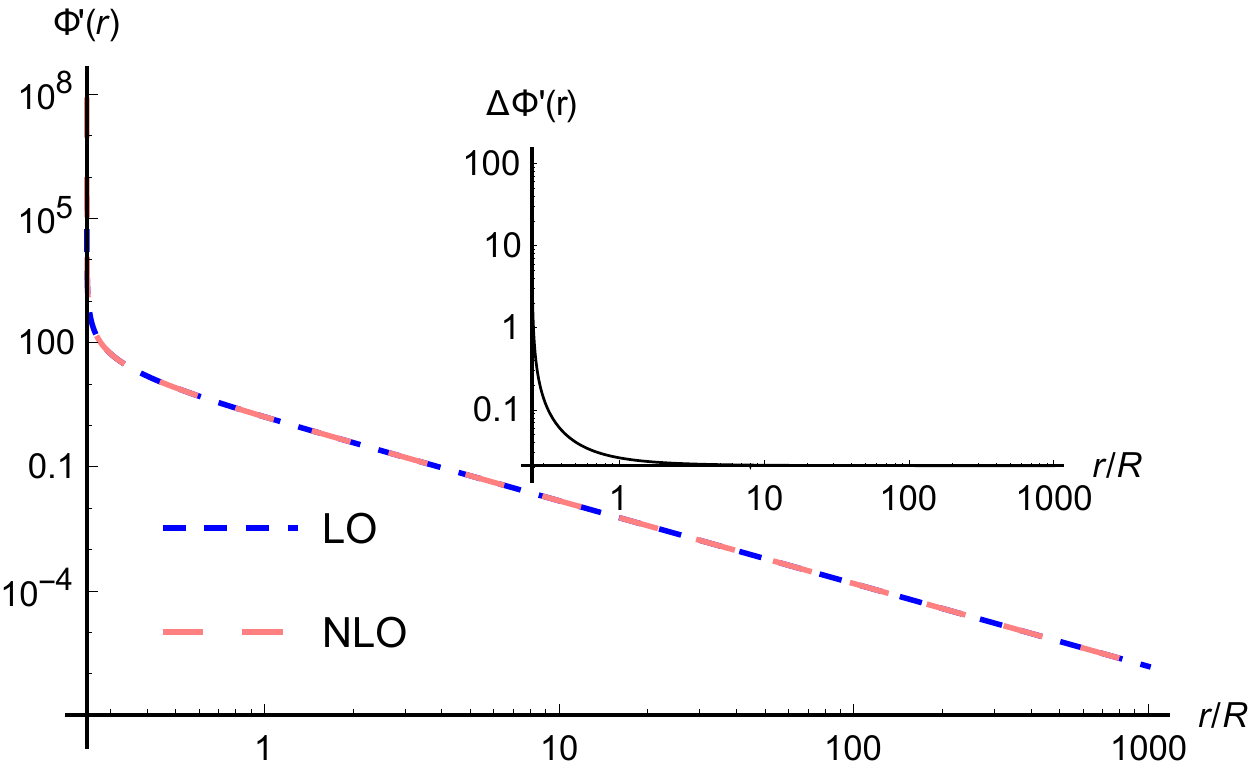}
\caption{The scalar hair $\Phi'(r)$ corresponding to the redshift function in Figure \ref{fig:metric_near_schw}. LO (NLO) stands for (next-to-)leading order which means the first (two) term(s) in Eq. \eqref{eq:schw_nu_asympt} is (are) considered at $r/R = 10^3$ where the integration is initiated. The inset shows the percent deviation between the LO and NLO curves.}
\label{fig:phi_near_schw}
\end{figure}

Now that the metric functions are derived from Eq. \eqref{eq:ttrr_vto_mi_J0}, we obtain the scalar field using Eq. \eqref{eq:rr_vto_mi}. Pragmatically speaking, in this step, a choice for the coupling $K_B$ must be made. We choose $K_B = 3$ for illustration but emphasize that any other choice works, except $K_B = 2$. This leads to the scalar hair shown in Figure \ref{fig:phi_near_schw} which corresponds to the LO and NLO solutions in Figure \ref{fig:metric_near_schw}.

We find that the scalar hair $\Phi'(r)$ on top of the nearly-Schwarzschild background features a divergence at the event horizon, which is also found in the hairy Schwarzschild solution (Eq. \eqref{eq:phi_hairy_schw}). This appears in both LO and NLO curves in Figure \ref{fig:phi_near_schw} where the inset shows the resolvable percent deviation between the two curves. It is interesting that $\Phi'(r)$ continuously drops as an inverse-square further away ($r/R \gg 1$) despite the metric showing resolvable deviation from the Schwarzschild black hole. Compared with the NLO case, the LO solution is expected to feature stronger deviations from the hairy Schwarzschild black hole presented in the previous section (Eqs. \eqref{eq:schw_nu}, \eqref{eq:schw_zeta}, and \eqref{eq:phi_hairy_schw}), as expected.

\begin{figure}[h!]
\center
\includegraphics[width = 0.8 \textwidth]{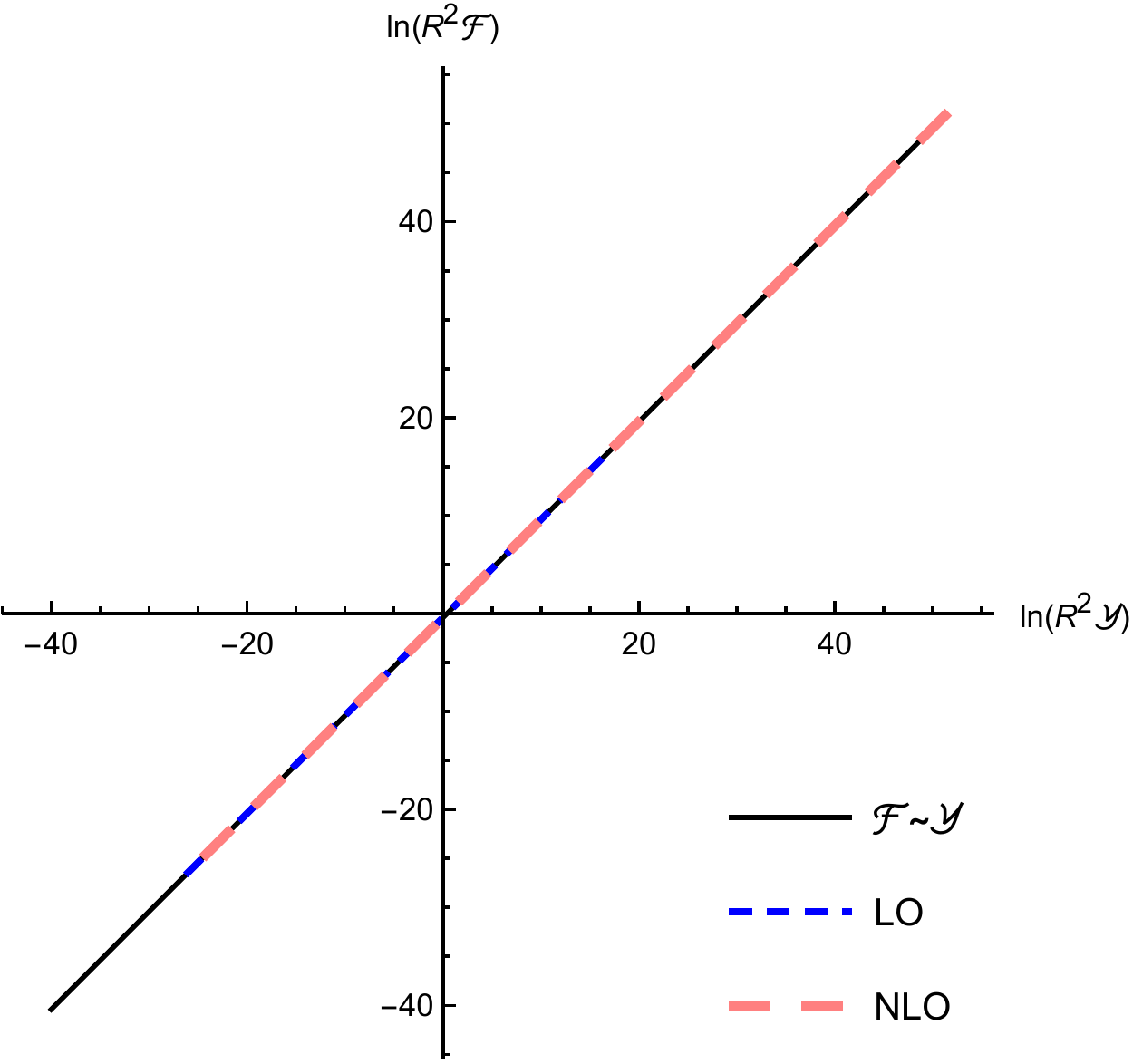}
\caption{The TeVeS potential ${F}$ corresponding to the nearly-Schwarzschild solution (Figures \ref{fig:metric_near_schw} and \ref{fig:phi_near_schw}). LO (NLO) stands for (next-to-)leading order which means the first (two) term(s) in Eq. \eqref{eq:schw_nu_asympt} is (are) considered at $r/R = 10^3$ where the integration is initiated.}
\label{fig:Fpot_near_schw}
\end{figure}

The designer approach finishes by solving the TeVeS potential ${F}$ using Eq. \eqref{eq:thetatheta_vto}. This leads to the potential shown in Figure \ref{fig:Fpot_near_schw} corresponding to the LO and NLO nearly-Schwarzschild solutions. The linear TeVeS potential (Eq. \eqref{eq:Fpot_hairy_schw}) supporting the hairy Schwarzschild black hole is also shown for $K_B = 3$.

It can be seen that the nearly-Schwarzschild solution in both the LO and NLO cases agree well with the linear TeVeS potential ${F} \sim {Y}$ (Eq. \eqref{eq:Fpot_hairy_schw}). Both cannot even be visually distinguished from Eq. \eqref{eq:Fpot_hairy_schw} except that they can only be resolved in a finite range where the numerical integration is evaluated. This is the designer approach. Deviations from the linear potential may be inspected at the asymptotically large distance ($r/R \gg 1$) where $R^2 {F} \ll 1$ and $R^2 {Y} \ll 1$.

Of course, the above nearly-Schwarzschild solutions were derived from necessary conditions and \textit{not} the modified Einstein field equations. Substituting these back to the field equations of the theory (Eqs. \eqref{eq:tt_vto}, \eqref{eq:rr_vto}, \eqref{eq:thetatheta_vto}, and \eqref{eq:sfe_vto}) is an important confirmation of the designing procedure. We were able to confirm this for the nearly-Schwarzschild solutions. Even more convincingly, this was the way in which we discovered the hairy Schwarzschild black hole solution (Eqs. \eqref{eq:schw_nu}, \eqref{eq:schw_zeta}, \eqref{eq:phi_hairy_schw}) in a linear TeVeS potential (Eq. \eqref{eq:Fpot_hairy_schw}). We end this subsection with a summary of the designer approach.

\subsection{A designer recipe for TeVeS}
\label{subsec:designer_recipe}

In this subsection, we would like to quickly summarize the recipe for the designer approach that helps us to find the aforementioned solutions. For a time-pointing vector field, the following recipe summarizes key steps for obtaining solutions in TeVeS:

\begin{enumerate}
    \item Specify either $\nu(r)$ or $\zeta(r)$, and solve for the other metric function using Eq. \eqref{eq:ttrr_vto_mi_J0}. Note: Eq. \eqref{eq:ttrr_vto_mi_J0} was derived with ${J} = 0$, and so must be used on the subsequent steps.
    \item Provided $(\nu, \zeta)$, solve for the hair $\Phi'(r)$ using Eq. \eqref{eq:rr_vto_mi}.
    \item Construct the coupling ${Y}(r) = e^{-\zeta(r)} \Phi'(r)^2$, and afterwards the TeVes potential ${F}[{Y}(r)]$ using Eq. \eqref{eq:thetatheta_vto}. The vector field and the multiplier can be obtained through Eqs. \eqref{eq:vector_vto} and \eqref{eq:multiplier_vto}.
    \item Confirm the solution $\left( \nu, \zeta, \Phi, {F} \right)$ by substituting it back into the field equations (Eqs. \eqref{eq:tt_vto}, \eqref{eq:rr_vto}, \eqref{eq:thetatheta_vto}, and \eqref{eq:sfe_vto}).
\end{enumerate}

We emphasize that the solutions constructed by means of this recipe lie on the hypersurface determined by the vanishing shift charge, i.e., ${J} = 0$, or in terms of the current, $S^r = 0$. Such solutions are often considered in scalar field theories as it keeps the norm of the shift current $S^a$ finite particularly at the event horizon of a black hole \cite{Babichev:2016fbg, Babichev:2017guv}. Here we consider the assumption ${J} = 0$ for illustrative purposes for the sake of simplicity; otherwise, the generalized version of Eq. \eqref{eq:ttrr_vto_mi_J0} with ${J} \neq 0$ is not presentable. The designing approach may be extended to solutions on ${J} \neq 0$ albeit with challenges.

\section{On a nonvanishing shift charge and a divergent scalar field \label{sec:on_vanishing_charge}}

In the previous section, we have restricted our attention only on zero shift charge solutions for simplicity. However, while we maintain the stance that the equations for a nonvanishing shift charge are not presentable, it is possible to proceed numerically \footnote{The interested reader may start with Eqs. \eqref{eq:tt_vto_mi} and \eqref{eq:rr_vto_mi} and take the similar steps from there.}. In this section, we include a nonvanishing shift charge and particularly describe the interplay between the black hole size, mass, and the shift charge.

\begin{figure}[h!]
\center
\includegraphics[width = 0.8 \textwidth]{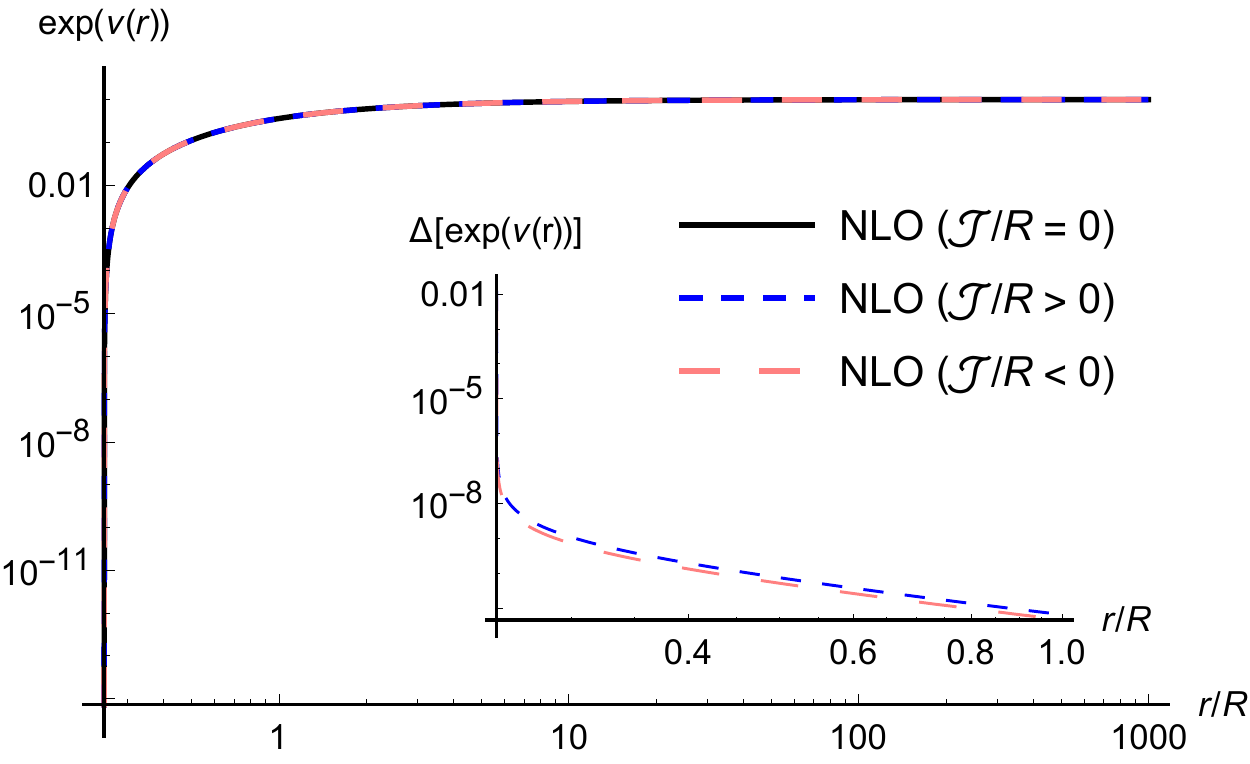}
\caption{The redshift function $\exp( \nu(r) )$ for nearly Schwarzschild solutions ($J = \mathcal{J}$) with $\mathcal{J}/R = 0$ and $\pm 10^{-1}$. We insert the NLO asymptotic condition Eq. \eqref{eq:schw_nu_asympt} at $r/R = 10^3$. The inset shows the absolute percent deviation of the ${J} \neq 0$ solutions compared to the NLO solution with ${J} = 0$.}
\label{fig:metric_NLO_J0pm}
\end{figure}

First of all, we find that the generalized version of Eq. \eqref{eq:ttrr_vto_mi_J0} after including a nonvanishing shift charge (${J} \neq 0$) continues to admit the Schwarzschild metric as an exact solution for arbitrary ${J}$ and $K_B$. We may formally write this down as 
\begin{equation}
0 = {H}_0[\nu, \zeta] + {J} {H}_1[\nu, \zeta; {J}, K_B]
\end{equation}
where ${H}_0$ is the functional defined in Eq. \eqref{eq:ttrr_vto_mi_J0} while the functional ${H}_1$ not only depends on the metric but also on the shift charge ${J}$ and the parameter $K_B$. Recalling that ${H}_0 \left[ \nu_\text{Schw}, \zeta_\text{Schw} \right] = 0$ on the Schwarzschild solution, we find ${H}_1 \left[ \nu_\text{Schw}, \zeta_\text{Schw}; {J}, K_B \right] = 0$ as well, independent of the shift charge. On the other hand, for nearly Schwarzschild solutions, we find that a nonzero shift charge increases the black hole size. Since the solutions are asymptotically flat, then this means that the shift charge alters the total mass of the black hole. We present numerical solutions with ${J}/R = \pm 10^{-1}$ in Figure \ref{fig:metric_NLO_J0pm} as an illustration.

To set a baseline, we recall that the horizon shift of the NLO solution, compared to the corresponding Schwarzschild black hole, is $\delta r_H/R = (r_H - r_\text{Schw})/R = -6 \times 10^{-12}$ with ${J} = 0$. Meanwhile, we find that $\delta r_H/R = 8 \times 10^{-10}$ for ${J}/R = 10^{-1}$ indicating that the black hole size has increased compared with the ${J}=0$ NLO solution. On the other hand, we find $\delta r_H/R = -5 \times 10^{-12}$ for ${J}/R = - 10^{-1} < 0$ reflecting a decrease in the black hole radius compared with the Schwarzschild one, while still larger than the ${J}=0$ case. We note that while there are caveats for larger values of the shift charge where the numerical integration breaks down unexpectedly, our investigation suggests that the above relation between the black hole size, mass, and shift charge holds for reasonably small values of $\| {J} \|$. We turn to the scalar field which presents a logarithmic divergence on the event horizon regardless of the shift charge.

We concur that the most controversial feature of the black holes in this paper is the logarithmic divergence of the scalar field on the event horizon. Such divergences were regarded as a physical obstacle to earlier incarnations of the model. In particular, in Bekenstein's TeVeS \cite{Bekenstein:2004ne}, similar hairy Schwarzschild black hole solutions emerge \cite{Giannios:2005es}. However, by looking at the physical frame, it became apparent that the logarithmic divergence of the background scalar hair implies inevitably a region near the horizon where $\Phi<0$, leading to a causality violation there. The difference between this and the present TeVeS is that the action \eqref{eq:rmond} is presented in the physical frame where matter is minimally coupled to the metric. This means that our black hole solutions which only feature tiny departures from the Schwarzschild geometry are presented in the frame of our experimental detectors and also are at least free of any further curvature singularities. Despite this, we are still yet to overcome technical strides to study the behavior of the gravitational perturbations on this background, which may reveal physical pathologies, instabilities of the solution.

\begin{figure}[h!]
\center
\includegraphics[width = 0.8 \textwidth]{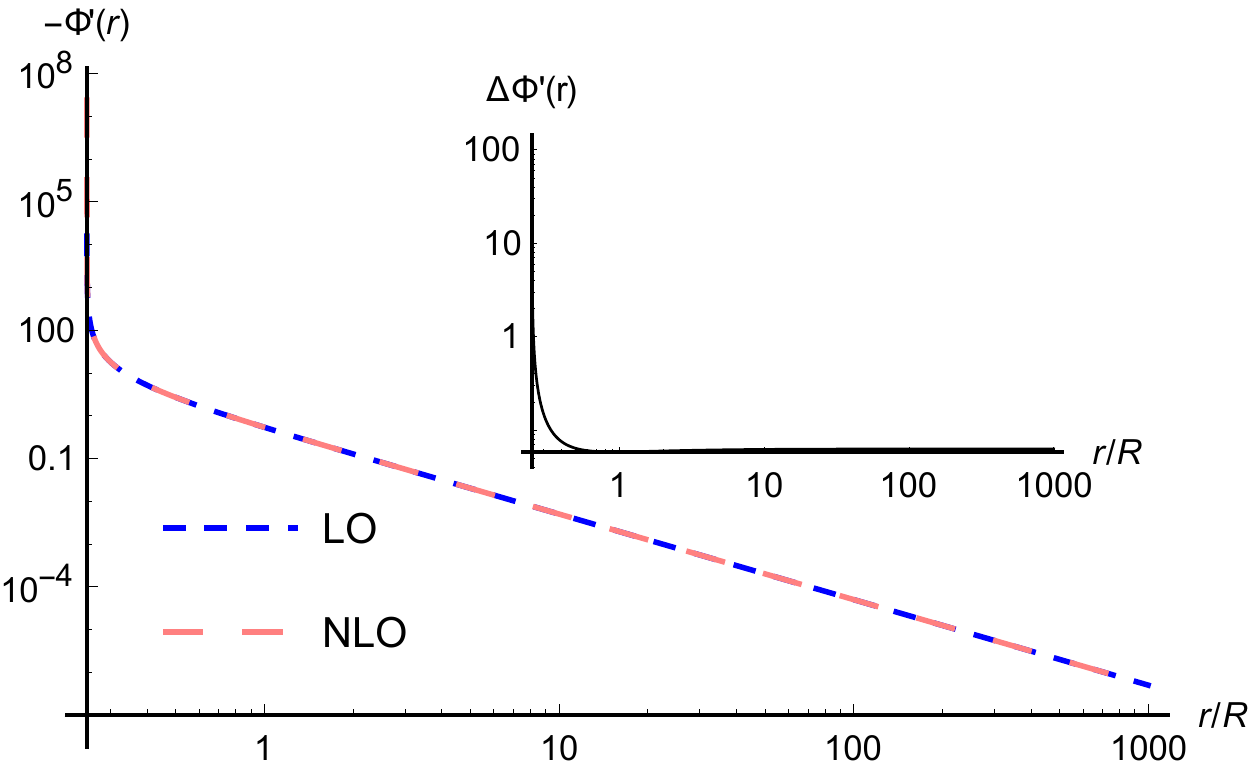}
\caption{The scalar hair $\Phi'(r) < 0$ ($K_B = 1$) corresponding to the redshift function in Figure \ref{fig:metric_near_schw}. LO (NLO) stands for (next-to-)leading order which means the first (two) term(s) in Eq. \eqref{eq:schw_nu_asympt} is (are) considered at $r/R = 10^3$ where the integration is initiated. The $y$-axis shows $-\Phi'(r) > 0$ in logarithmic scale. The inset shows the percent deviation between the LO and NLO curves.}
\label{fig:phi_K1}
\end{figure}

We may also mention that even if the causality violation recognized in Bekenstein's TeVeS holds, this pathology is tied to the scalar field being negative at the vicinity of the event horizon. This could be completely avoided in the present model by choosing $K_B \in (0, 2)$ in such a way that, close to the event horizon $r = R_H$, $\Phi' \sim a / \left( r - R_H \right) < 0$, or rather that $\Phi \sim a \ln \left( r - R_H \right) > 0$, for some constant $a = K_B/\left(K_B - 2\right) < 0$. This is shown explicitly in Figure \ref{fig:phi_K1}. Note also that the choice $K_B \in (0, 2)$ was made in Ref.~\cite{Skordis:2020eui} and was shown to be consistent with cosmological observations.

We note that $-\Phi'(r) > 0$ is plotted on the $y$-axis in logarithmic scale in Figure \ref{fig:phi_K1}. This is to make clear that $\Phi'(r) < 0$ for $K_B = 1$ which lets the overall solution overcome causality violation through $\Phi(r \sim R_H + \epsilon) > 0$ as detailed in Ref. \cite{Giannios:2005es}. In general, $\Phi'(r) < 0$ for $K_B \in (0, 2)$ and $\Phi'(r) > 0$ otherwise. Understandably, this was an excursion and we remind that our hairy Schwarzschild solution is presented in the physical frame. In the following section, we see whether granting the vector field a physical orientation could alleviate the unwanted scalar field divergence in this model.

\bigskip

\section{{Vector field with nonvanishing radial components}}
\label{sec:time_space_vector}

The hairy Schwarzschild solution presented in section \ref{subsec:analytical} is an elegant, analytical solution representing the strong gravity regime in this TeVeS model. The linear coupling, ${F} \sim {Y}$, also appears to be most natural from an effective field theory perspective. However, it does come with the unpleasant feature of the scalar hair (Eq. \eqref{eq:phi_hairy_schw}) diverging on the event horizon of the black hole. In this section, we would like to investigate whether this concern is really a physical pathology or not, and whether it can be alleviated, in particular, when the vector field has a non-zero projection along the spatial direction.

{We clarify, as is well known, that the components of a vector are gauge dependent, in the sense that one could always find a coordinate system in which a vector field has vanishing components in some directions. However, we have let go of this generality upon working in the isotropic coordinates. While a coordinate system surely exists where the vector field's spatial components vanish, the transformation from isotropic coordinates to this new coordinate system may not be pretty, if even `pragmatically' achievable. Just the same, this coordinate system is of no concern to us in this section, but rather we focus on utilizing what remains of the vector field's degrees of freedom in isotropic coordinates to attempt to clear the fields' divergence on the Schwarzschild black hole event horizon.} 

We consider the linear TeVeS potential (Eq. \eqref{eq:Fpot_hairy_schw}). Then, with ${V} \neq 0$, it can be shown that the generalization to the hairy Schwarzschild black hole is given by Eqs. \eqref{eq:schw_nu} and \eqref{eq:schw_zeta} together with the scalar hair
\begin{equation}
\label{eq:phi_schw_hairy_genvec}
    \Phi'(r) = \dfrac{8 K_B}{(K_B-2)} \dfrac{{H}[{V}(r)]}{(16 r^2-R^2) \left(256 r^4 {V}(r)^2+(4 r+R)^4\right)} \,,
\end{equation}
vector hair
\begin{equation}
\label{eq:Uvec_schw_hairy_genvec}
    {U}(r) = \frac{(4 r-R) \sqrt{256 r^4 {V}(r)^2+(4 r+R)^4}}{(4 r+R)^3}\,,
\end{equation}
where the functional ${H}[{V}(r)]$ is given by
\begin{equation}
\label{eq:Hfunc_schw_hairy}
\begin{split}
    {H}[{V}(r)] = 32 r^3 {V}(r) \bigg( 
    & r \left(16 r^2-R^2\right) {V}'(r) \\
    & +2 R (8 r-R) {V}(r)\bigg)+R (4 r+R)^4 \,.
\end{split}
\end{equation}
The multiplier is given by
\begin{equation}
\label{eq:multiplier_schw_hairy}
\begin{split}
    \lambda = \ & \bigg( 16384 K_B r^4 \bigg( 4 r^2 \bigg(r (4 r-R) (4 r+R)^5 {V}(r) \\
    & \phantom{gg} \times \left(r \left(16 r^2-R^2\right) {V}''(r)+8 \left(4 r^2+6 r R-R^2\right) {V}'(r)\right) \\
    & \phantom{gg} +2 R^2 \left(208 r^2-48 r R+3 R^2\right) (4 r+R)^4 {V}(r)^2 \\
    & \phantom{gg} +r^2 (R-4 r)^2 (4 r+R)^6 {V}'(r)^2 \\
    & \phantom{gg} +256 r^5 \left(16 r^2-R^2\right) {V}(r)^3 \\
    & \phantom{gg} \times \left(r \left(16 r^2-R^2\right) {V}''(r)+4 \left(8 r^2+8 r R-R^2\right) {V}'(r)\right) \\
    & \phantom{gg} +512 r^4 R^2 \left(144 r^2-32 r R+R^2\right) {V}(r)^4\bigg) \\
    & \phantom{gg} +R^2 (4 r+R)^8 \bigg) \bigg) \bigg/ (R-4 r)^2 (4 r+R)^6 \\
    & \phantom{ggggggggggggggggggggg} \times \left(256 r^4 {V}(r)^2+(4 r+R)^4\right)^2 \,.
\end{split}
\end{equation}
The metric (Eqs. \eqref{eq:schw_nu} and \eqref{eq:schw_zeta}), scalar (Eq. \eqref{eq:phi_schw_hairy_genvec}), and vector fields (Eq. \eqref{eq:Uvec_schw_hairy_genvec}) can be confirmed to be a solution of the field equations (Eqs. \eqref{eq:Ett}, \eqref{eq:Etr}, \eqref{eq:Err}, \eqref{eq:Ethetatheta}, \eqref{eq:sfe}, \eqref{eq:vfe_t}, and \eqref{eq:vfe_r}) with the linear TeVeS potential (Eq. \eqref{eq:Fpot_hairy_schw}). This general solution reduces to the time-pointing vector counterparts (Eqs. \eqref{eq:phi_hairy_schw} and \eqref{eq:multiplier_schw_hairy}) in the limit ${V} \rightarrow 0$.

The takeaway from this general solution is that this holds for an arbitrary, unspecified function ${V}(r)$ which corresponds to how the vector field is projected along the spatial direction. This degree of freedom can function to overcome the divergence of the scalar hair on the event horizon. We demonstrate this in general and also explicitly below.

Eq. \eqref{eq:phi_schw_hairy_genvec} implies $\Phi'(r) \sim 1/(r- R)$ for constant ${V}$ (including the limit ${V}(r) = 0$). To take care of this divergence, we take advantage of choosing the spatial degree of freedom of the vector field ${V}(r)$ and orient it such that
\begin{equation}
\label{eq:regular_general}
    {H}[{V}(r)] = \left( 4r - R \right) {W}(r)
\end{equation}
where ${W}(r)$ is a function chosen to cure the divergence. In this case, the scalar field becomes
\begin{equation}
    \Phi'(r) = \dfrac{8 K_B}{(K_B-2)} \dfrac{{W}(r)}{(4r + R) \left(256 r^4 {V}(r)^2+(4 r+R)^4\right)}
\end{equation}
where the spatial component of the vector field is given by
\begin{equation}
    {V}(r) = {H}^{-1}\left[ \left( 4r - R \right) {W}(r) \right] \,.
\end{equation}
The factor causing the scalar hair $\Phi'(r)$ to diverge on the event horizon is cancelled out. This leaves the vector field components with a horizon divergence which could be cured by orienting it (or rather choosing its initial conditions) appropriately.

We illustrate this for a particular case with
\begin{equation}
\label{eq:regular}
    {H}[{V}(r)] = \eta (4r - R)\,,
\end{equation}
where $\eta$ is a constant. The exact solution to Eq. \eqref{eq:regular} could be written in closed form. However, this analytical expression is not too informative. Instead, its leading order asymptotic term at the horizon is given by
\begin{equation}
\label{eq:vr_schw_hor}
    {V}\left(r\sim R/4\right) \sim \dfrac{\sqrt{240(1024 \xi R^6- (\eta R^2/60) +4 R^6)}}{4 \sqrt{15} R^2 \left(r- (R/4)\right)}\,,
\end{equation}
where $\xi$ is an integration constant. The constants $(\eta, \xi)$ describe the spatial orientation of the vector field. With this, the asymptotic behavior of the scalar field and the time component of the vector field become
\begin{equation}
\label{eq:phi_schw_hor}
    \Phi'(r \sim R/4) \sim \frac{15 \eta K_B (R-4 r)^2}{(K_B-2) R^3 \left(240 \left(256 \xi + 1\right) R^4-\eta \right)}\,,
\end{equation}
and
\begin{equation}
\label{eq:vt_schw_hor}
    {U}(r \sim R/4) \sim \dfrac{\sqrt{1024 \xi R^6- (\eta R^2/60) +4 R^6}}{2 R^3} \,.
\end{equation}
It is clear in Eq. \eqref{eq:phi_schw_hor} that the concern regarding the divergence of the scalar field on the event horizon has gone away by choosing the spatial orientation of the vector according to Eq. \eqref{eq:regular}. This turned the spatial component of the vector field (Eq. \eqref{eq:vr_schw_hor}) to be singular on the event horizon. We make use of the integration constant $\xi$ to then completely fix the spatial orientation of the vector field to overcome this, i.e., choose $\eta$ such that $1024 \xi R^6- (\eta R^2/60) +4 R^6 = 0$. We then finally obtain a hairy Schwarzschild black hole with scalar and vector fields:
\begin{equation}
\label{eq:phi_schw_divfree}
\begin{split}
    \Phi'(r) = & \dfrac{1920 K_B R^4}{(K_B-2) \left(16 r^2-R^2 \right) } \times \dfrac{1}{64 r^3+144 r^2 R+156 r R^2+111 R^3} \,,
\end{split}
\end{equation}
\begin{equation}
\label{eq:Uvec_schw_divfree}
\begin{split}
    {U}(r) = & \dfrac{(4 r-R)^{3/2}}{4 \sqrt{15} R^2 (4 r+R)^3} \sqrt{\eta  \left(64 r^3+144 r^2 R+156 r R^2+111 R^3\right)} \,,
\end{split}
\end{equation}
and
\begin{equation}
\label{eq:Vvec_schw_divfree}
\begin{split}
    {V}(r) = \bigg( & -16 R^4 (4 r+R)^4 -\frac{1}{15} \eta  (R-4 r) \\
    & \times \left(64 r^3+144 r^2 R+156 r R^2+111 R^3\right) \bigg)^{1/2} \bigg/ \left(64 r^2 R^2\right) \,.
\end{split}
\end{equation}
These surround the Schwarzschild black hole (Eqs. \eqref{eq:schw_nu} and \eqref{eq:schw_zeta}) in TeVeS. These hairs (Eqs. \eqref{eq:phi_schw_divfree}, \eqref{eq:Uvec_schw_divfree}, and \eqref{eq:Vvec_schw_divfree}) are clearly well-defined at infinity as
\begin{equation}
    \Phi'(r \sim \infty) \sim \dfrac{15 K_B R^4}{8 (K_B-2) r^5} \phantom{\dfrac{\dfrac{1}{1}}{\dfrac{1}{1}}}\,,
\end{equation}
\begin{equation}
    {U}(r\sim \infty) \sim \dfrac{\sqrt{\gamma }}{4 \sqrt{15} R^2}-\dfrac{\sqrt{15} \sqrt{\gamma } R^2}{128 r^4} \,,
\end{equation}
and
\begin{equation}
    {V}(r\sim \infty) \sim \dfrac{\sqrt{ \gamma -240 R^4 }}{4 \sqrt{15} R^2}-\dfrac{120 R^4-\gamma }{4 r \left(\sqrt{15} R \sqrt{\gamma -240 R^4}\right)} \,.
\end{equation}
However, by choosing the vector to be regular, it should be noted that the divergence on the event horizon has come back once again to the scalar field (Eq. \eqref{eq:phi_schw_divfree}). Our examinations suggest that only either one the scalar or the vector field can be regular. We suspect this could be due to the Schwarzschild background, not being able to accommodate regular hair. Constraints on black holes by gravitational waves suggest that GR, and so its solution, the Schwarzschild black hole, is an excellent description of nonspinning compact remnants from postmerging binaries. Regardless of the compatibility of the Schwarzschild geometry in a TeVeS representation of RMoND, this implies that the Schwarzschild solution can be regarded as a bona fide, first approximation to realistic, astrophysical black holes. Fields standing on the Schwarzschild geometry should then leave traces which could be picked up by observations.

{It is fair to ask if the field divergences on the event horizon we find are a consequence of the coordinate choice. This is however not the case at all since the fields are static. In particular, transforming to Eddington-Finkelstein coordinates $(v = t - r_*(x), x)$, where $x = r\left(1 + R/4r\right)^2$, we find that the scalar field diverges logarithmically, $\Phi'(x) \sim 1/(x - R)$, so does the gauge invariant kinetic term $Y \sim 1/(x - R)$. This is indicative of singular sources of the fields at the event horizon, and prevents us from probing the behavior of the fields at the interior of the black hole.}

{We classify the scalar and vector charges of the Schwarzschild and nearly Schwarzschild black holes presented in this work to be of secondary type \cite{Babichev:2016rlq}, in the sense that these `hairs' are controlled by fixed mass scales of the theory rather than by arbitrary constants. Regardless, the divergence of these charges on the event horizon positions us to question the physicality of the solution. In Ref. \cite{R:2022cwe} for example it was shown that analogous divergent scalar fields make the black holes difficult to be formed through gravitational collapse. While this may also hold here, the gravitational couplings in TeVeS are richer and more complex, requiring an independent study of the gravitational collapse with scalar and vector charges, which would make by itself an interesting work.}

\section{Outlook}
\label{sec:outlook}

We have presented spherically symmetric solutions in the TeVeS theory (Eq. \eqref{eq:rmond}) representative of the strong gravity regime. Our results have shown that nontrivial fields do exist on top of Schwarzschild and nearly-Schwarzschild black holes; however, these come with an undesirable diverging hair (from either the scalar or vector fields) on the event horizon.

Schwarzschild black holes are of course by no means necessarily compatible with any alternative gravity. An unwanted divergence for one solution (whether pathological or not) may be alleviated when considering spherically symmetric spacetimes beyond the Schwarzschild solution. However, gravitational wave constraints on the post-Newtonian parameters show only an ever increasing support for GR \cite{LIGOScientific:2021sio}. In the nonrotating limit, real, astrophysical black holes must therefore look like a Schwarzschild black hole in a first approximation. Any theory or phenomenological model should reflect this. It is in this spirit that we opted to work on isotropic coordinates where the post-Newtonian parameters are transparent should a departure from the Schwarzschild geometry exist. The behavior of nontrivial fields on a Schwarzschild black hole translate to whether a theory may be phenomenologically useful or even physically valid. In TeVeS, we found that a hairy Schwarzschild solution can be accommodated by means of a linear potential. The diverging scalar or vector fields that come with this therefore pose a challenge for the model. This remains so even when the fields are dark, uncoupled to visible matter, and become singular only on the event horizon.

Any physical consequence of the divergence may be hidden in this regard since matter (and light) simply falls on the geodesics of the Schwarzschild metric. Another conclusive view to see whether the divergence is pathological relies on an analysis of the gravitational perturbations to determine whether this background can hold its own against small disturbances. The logarithmic behavior of the background fields on the event horizon would alter the boundary conditions of the perturbations, inevitably largely influencing the stability of the system. If the solution turns out to be unstable, then this puts the TeVeS framework less favorable from both a physical and observational standpoint. However, this comes with serious technical obstacles -- a derivation of the modified master equation for the gravitational perturbations, an analysis of the asymptotic behavior of waves particularly at the horizon where the background fields diverge, a precise numerical scheme for solving higher dimensional coupled scalar-vector-tensor wave equations, etc. -- which put it beyond the scope of the present paper. Nonetheless, we have made remarkable progress in the gravitational perturbations around general relativistic black holes in the TeVeS \eqref{eq:rmond} which we hope to discuss in a different paper \cite{rcb_cyc_in_prep}.

We emphasize that our results simply mean that nontrivial fields in the strong gravity regime of RMoND and generally TeVeS need further assessment for physical pathologies. Gravitational lensing around black holes for instance may be considered to reveal nontrivial degrees of freedom \cite{Zhang:2021ygh}. Alternatively, this may be taken as a hint that the fields must be trivial on a black hole. Observational signatures of the dark fields then emerge by disturbing the perturbations on top of the black hole, leading to a wave of exciting phenomena such as anomalous quasinormal modes \cite{Tattersall:2018nve, Lagos:2020oek} and tilting of the waveforms due to a coupling between the dark fields \cite{Bernardo:2021vsj}. We shall discuss this elsewhere.

The existence of black holes is now an important test of any theory, given the vast astronomical observations (e.g., gravitational waves, very long baseline interferometry) that support their astrophysical relevance. Whether black hole spacetimes in the new TeVeS pass state-of-the-art black hole observations was the open question that motivated this work. Our paper adds to MoND phenomenology by examining not just the weak field limit but most importantly the strong gravity regime of its TeVeS realization, and showing that black holes can exist in the theory, thereby increasing the support for it. Furthermore, the solutions we found are consistent with the Newtonian limit of the post-Newtonian expansion, and so measurable departures from one of the parameterized post-Newtonian parameters $\gamma$ and $\beta$ disfavor these solutions, and indirectly the theory. Witnessing a tightening of the existing parameterized post-Newtonian parameter constraints have thus become more exciting for MoND in light of our results.

\section*{Declarations}


\begin{itemize}
\item Data availability: Data sharing not applicable to this article as no datasets were generated or analysed during the current study.
\item Competing interests: The authors have no competing interests to declare that are relevant to the content of this article.
\item Code availability: A Mathematica notebook for the derivation of the results of this work can be downloaded from \href{https://github.com/reggiebernardo/notebooks}{GitHub} \cite{reggie_bernardo_4810864}.
\end{itemize}







\begin{appendices}






\section{Variation toolkit}
\label{sec:variation}

We present metric functional variations ($\delta g^{ab}$) of relevant quantities that can be used to derive the stress-energy tensor (Eq. \eqref{eq:set_mond}) containing the MoND degrees of freedom. The familiar curvature-related ones are
\begin{equation}
    \delta  \sqrt{-g} = - \dfrac{1}{2} \sqrt{-g} g_{ab} \delta g^{ab} \,,
\end{equation}
\begin{equation}
\delta \left( \sqrt{-g} R \right) = \sqrt{-g} G_{ab} \delta g^{ab} + \text{ boundary term} \,,
\end{equation}
and
\begin{equation}
\begin{split}
    \delta \Gamma^c_{ab} = - \dfrac{1}{2} \bigg[ g_{da} \nabla_b \left( \delta g^{dc} \right)
    & + g_{db} \nabla_a \left( \delta g^{dc} \right) - g_{ai} g_{bj} \nabla^c \left( \delta g^{ij} \right) \bigg] \,.
\end{split}
\end{equation}
On the other hand, variations of the various couplings between the vector and scalar degrees of freedom are given by
\begin{equation}
    \delta \left( F^{ab} F_{ab} \right) = 2 F_a^{\ c} F_{b c} \delta g^{ab} \,,
\end{equation}
\begin{equation}
    \delta \left( \nabla_\alpha A_\nu \right) = A_{(\beta} g_{\gamma) (\alpha} \nabla_{\nu)} \left( \delta g^{\gamma \beta} \right) - \dfrac{1}{2} g_{\alpha \gamma} g_{\lambda \nu} A^\beta \nabla_\beta \left( \delta g^{\gamma \lambda} \right) \,,
\end{equation}
\begin{equation}
\begin{split}
    \delta \left( J^a \nabla_a \phi \right) = & \bigg[ \left( \nabla_c \phi \right) A_{(a} \left( \nabla_{b)} A^c \right) + A^c \left( \nabla_c A_{(a} \right) \left( \nabla_{b)} \phi \right) \bigg] \delta g^{ab} \\
    & \phantom{gg} - g_{a(c} \nabla_{d)} \left[ A_b A^c \left( \nabla^d \phi \right) \right] \delta g^{ab} + \dfrac{1}{2} \nabla_c \left( A^c A_a \left( \nabla_b \phi \right) \right) \delta g^{ab} \\
    & \phantom{gg} + \text{ boundary terms, e.g., } \nabla_c \left( \cdots \right) \,,
\end{split}
\end{equation}
\begin{equation}
    \delta {Y} = \left[ \nabla_a \nabla_b \phi + 2 \left( A^c \nabla_c \phi \right) A_{(a} \nabla_{b)} \phi \right] \delta g^{ab} \,,
\end{equation}
\begin{equation}
\delta {Q} = \left( A_{(a} \nabla_{b)} \phi \right) \delta g^{ab}	\,,
\end{equation}
and
\begin{equation}
    \delta \left( A_a A^a \right) = A_a A_b \delta g^{ab} \,.
\end{equation}

\section{Explicit field equation terms in isotropic coordinates}
\label{sec:explicit_field_equations}

The explicit terms appearing in the modified Einstein equation components in isotropic coordinates are
\begin{equation}
\label{eq:Ett}
\begin{split}
E_{tt} = \ & e^{-\zeta } (K_B-2) {U}^2 \Phi '' - e^{\nu -\zeta } \zeta '' + \dfrac{e^{-2 \zeta }}{4 r} \bigg[ -2 r e^{2 \zeta +\nu } {F}({Y},{Q}) \\
& -8 e^{\zeta +\nu } \zeta '-r e^{\zeta +\nu } \left(\zeta '\right)^2+2 K_B r e^{\zeta +\nu } \left(\Phi '\right)^2 -4 r e^{\zeta +\nu } \left(\Phi '\right)^2 \\
& +2 e^{\zeta } K_B r {U}^2 \zeta ' \Phi '-4 e^{\zeta } r {U}^2 \zeta ' \Phi '-2 e^{\zeta } r \lambda  \left(e^{\zeta } {U}^2+e^{\nu } \left(e^{\zeta }+{V}^2\right)\right) \\
& +8 e^{\zeta } K_B r {U} {U}' \Phi ' -2 e^{\zeta } K_B r \left({U}'\right)^2 -16 e^{\zeta } r {U} {U}' \Phi ' \\
& +2 K_B e^{\nu } r {V}^2 \zeta ' \Phi '-4 e^{\nu } r {V}^2 \zeta ' \Phi '+2 K_B e^{\nu } r {V}^2 \left(\Phi '\right)^2-4 e^{\nu } r {V}^2 \left(\Phi '\right)^2 \\
& -4 K_B e^{\nu } r {V} {V}' \Phi ' +8 e^{\nu } r {V} {V}' \Phi '+8 e^{\zeta } K_B {U}^2 \Phi '-16 e^{\zeta } {U}^2 \Phi ' \bigg] \,,
\end{split}
\end{equation} 
and
\begin{equation}
\label{eq:Err}
\begin{split}
E_{rr} = \ & e^{-\zeta } (K_B-2) {V}^2 \Phi '' + \dfrac{1}{4} \bigg[ -8 e^{-\zeta } {V}^2 \left(\Phi '\right)^2 {F}_{{Y}} -4 {V} \Phi ' {F}_{{Q}} \\
& -4 \left(\Phi '\right)^2 {F}_{{Y}} +2 e^{\zeta } {F}({Y},{Q}) +2 \zeta ' \nu '+\left(\zeta '\right)^2+2 K_B \left(\Phi '\right)^2 \\
& +\frac{4 \zeta '}{r}+\frac{4 \nu '}{r}+\frac{8 e^{-\zeta } K_B {V}^2 \Phi '}{r}-\frac{16 e^{-\zeta } {V}^2 \Phi '}{r} -2 K_B e^{-\nu } {U}^2 \nu ' \Phi ' \\
& +4 e^{-\nu } {U}^2 \nu ' \Phi '+\lambda  \left(2 e^{\zeta }-2 {U}^2 e^{\zeta -\nu }-2 {V}^2\right) +2 K_B e^{-\nu } \left({U}'\right)^2 \\
& +4 e^{-\zeta } K_B {V}^2 \zeta ' \Phi ' -8 e^{-\zeta } {V}^2 \zeta ' \Phi '+2 e^{-\zeta } K_B {V}^2 \nu ' \Phi ' \\
& +6 e^{-\zeta } K_B {V}^2 \left(\Phi '\right)^2 -4 e^{-\zeta } {V}^2 \nu ' \Phi ' -12 e^{-\zeta } {V}^2 \left(\Phi '\right)^2 \\
& -4 e^{-\zeta } K_B {V} {V}' \Phi '+8 e^{-\zeta } {V} {V}' \Phi '-4 \left(\Phi '\right)^2 \bigg] \,,
\end{split}
\end{equation}
\begin{equation}
\label{eq:Etr}
\begin{split}
E_{tr} = \ & (K_B-2) {U} {V} \Phi ''-\dfrac{{U}}{2 r} \bigg[{V} \bigg(\Phi ' \bigg(2 r \Phi ' {F}_{{Y}}-4 K_B \\
& -(K_B-2) r (\zeta' + \nu') -2 (K_B - 2) r \Phi '+8\bigg)+2 e^{\zeta } r \lambda \bigg)+e^{\zeta } r \Phi ' {F}_{{Q}} \bigg] \,,
\end{split}
\end{equation}
and
\begin{equation}
\label{eq:Ethetatheta}
\begin{split}
E_{\theta\theta} = \
& \frac{r^2 \zeta ''}{2}+\frac{r^2 \nu ''}{2} + \frac{1}{4} r e^{-\zeta -\nu }
\bigg[ 2 r e^{2 \zeta +\nu } {F}({Y},{Q})+2 e^{\zeta +\nu } \zeta '+2 e^{\zeta +\nu } \nu ' \\
& -2 (K_B - 2) r e^{\zeta +\nu } \left(\Phi '\right)^2 +r e^{\zeta +\nu } \left(\nu '\right)^2 +2 e^{\zeta } (K_B - 2) r {U}^2 \nu ' \Phi ' \\
& +2 e^{\zeta } r \lambda  \left(e^{\nu } \left(e^{\zeta }+{V}^2\right)-e^{\zeta } {U}^2\right)-2 e^{\zeta } K_B r \left({U}'\right)^2 - 2 (K_B - 2) e^{\nu } r {V}^2 \zeta ' \Phi ' \\
& -2 (K_B - 2) e^{\nu } r {V}^2 \left(\Phi '\right)^2 +4 (K_B - 2) e^{\nu } r {V} {V}' \Phi ' \bigg] \,.
\end{split}
\end{equation}

\section{The curvature scalars in isotropic coordinates \label{sec:curvature_scalars}}

The Ricci $\mathcal{R}$ and Kretschmann $\mathcal{K}$ scalars in isotropic coordinates \eqref{eq:iso_coords} are given by
\begin{equation}
\label{eq:ricci}
\begin{split}
    \mathcal{R} = -\frac{e^{-\zeta (r)}}{2 r} \bigg(
    & 2 r \left(2 \zeta ''(r)+\nu ''(r)\right) + \zeta '(r) \left(r \nu '(r)+8\right) \\
    & \ \ + r \zeta '(r)^2 +r \nu '(r)^2+4 \nu '(r) \bigg)
\end{split}
\end{equation}
and
\begin{equation}
\label{eq:kretschmann}
\begin{split}
    \mathcal{K} = \frac{e^{-2 \zeta (r)}}{4 r^2} \bigg(
    & 4 r^2 \left(2 \zeta ''(r)^2+\nu ''(r)^2\right) +3 \zeta '(r)^2 \left(r^2 \nu '(r)^2+8\right) \\
    & \ \ + r^2 \zeta '(r)^4 + r^2 \nu '(r)^4+4 \nu '(r)^2 \left(r^2 \nu ''(r)+2\right) \\
    & \ \ + 8 r \zeta '(r)^3-2 r \zeta '(r) \bigg(-8 \zeta ''(r) \\
    & \ \ +r \nu '(r)^3-4 \nu '(r)^2+2 r \nu '(r) \nu ''(r) \bigg) \bigg) \,.
\end{split}
\end{equation}
For a Schwarzschild black hole (Eqs. \eqref{eq:schw_nu} and \eqref{eq:schw_zeta}), these can be shown to be $\mathcal{R} = 0$ and $\mathcal{K} = 12 R^2/ \left( r^6 \left( 1 + (R/4r) \right)^{12} \right)$ as expected.

\section{Constrained vector field solutions}
\label{sec:unit_v_sol}

In this appendix, we examine the branch $\Phi'(r) = 0$ with $K_B \neq 0$. In this case, the $r$-component of the vector field equation as well as the $tr$-component of the metric field equation reduce to ${V}(r) \lambda(r) = 0$, i.e., either the multiplier $\lambda$ must be zero or that the vector field should be time-pointing. Also, due to the term $\sim J^a \nabla_a \phi$ in the action, despite having a trivial scalar field, the scalar field equation only becomes an additional nontrivial constraint on the vector field components. Thus, we take $K_B = 2$; otherwise, the dynamical system becomes over-constrained. We consider separately $\lambda = 0$ and ${V}(r) = 0$ branches below.

We thread down the $\lambda = 0$ branch together with ${F}(0, 0) = 0$ (asymptotically flat condition). The metric equation becomes
\begin{equation}
\label{eq:tt_lambda0}
    -e^{\nu} \left(4 r \zeta ''+r \zeta ^{\prime 2}+8 \zeta '\right)-2 K_B r {U}^{\prime 2} = 0 \,,
\end{equation}
\begin{equation}
\label{eq:rr_lambda0}
    2 \zeta ' \left(r \nu '+2\right)+r \zeta^{\prime 2}+4 \nu '+2 K_B r e^{-\nu (r)} {U}^{\prime 2} = 0\,,
\end{equation}
while the time component of the vector field equation becomes
\begin{equation}
\label{eq:vfe_lambda0}
    2 r {U}''+{U}' \left(r \zeta '-r \nu '+4\right) = 0 \,.
\end{equation}
In addition, the unit-vector constraint is given by
\begin{equation}
\label{eq:unit_vec_lambda0}
    e^{\nu} \left(e^{\zeta}+{V}^2\right)=e^{\zeta } {U}^2 \,.
\end{equation}
Eqs. \eqref{eq:tt_lambda0}, \eqref{eq:rr_lambda0}, \eqref{eq:vfe_lambda0}, and \eqref{eq:unit_vec_lambda0} can be solved for the four unknowns $\left( \nu. \zeta, {U}, {V} \right)$. To obtain this general solution, we solve Eq. \eqref{eq:vfe_lambda0} for ${U}$:
\begin{equation}
\label{eq:U_lambda0}
    {U}(r) = c_1 \int^r \dfrac{ \exp \left( \left( \nu (x)- \zeta (x) \right)/2 \right) }{x^2} \, dx + c_2\,,
\end{equation}
where $c_1$ and $c_2$ are integration constants. Substituting Eq. \eqref{eq:U_lambda0} into \eqref{eq:unit_vec_lambda0} then determines the spatial component of the vector field in terms of the metric functions $\left( \nu, \zeta \right)$. A coupled system for the metric functions finally emerges by substituting Eq. \eqref{eq:U_lambda0} into Eqs. \eqref{eq:tt_lambda0} and \eqref{eq:rr_lambda0}:
\begin{equation}
\label{eq:tt_final_lambda0}
    2 c_1^2 K_B+4 r^4 e^{\zeta} \zeta ''+r^4 e^{\zeta} \zeta ^{\prime 2}+8 r^3 e^{\zeta} \zeta ' = 0\,,
\end{equation}
and
\begin{equation}
\label{eq:rr_final_lambda0}
    \dfrac{2 c_1^2 K_B e^{-\zeta }}{r^3}+2 \zeta ' \left(r \nu '+2\right)+r \zeta ^{\prime 2}+4 \nu ' = 0 \,.
\end{equation}
An exact analytical solution to Eqs. \eqref{eq:tt_final_lambda0} and \eqref{eq:rr_final_lambda0} for $c_1 \neq 0$ cannot be obtained. Nonetheless, it is easy to confirm that for $c_1 = 0$, $\nu$ and $\zeta$ reduces to the Schwarzschild solution (Eqs. \eqref{eq:schw_nu} and \eqref{eq:schw_zeta}). This suggests that a hairy black hole emerges for $c_1 \neq 0$, in which case $c_1$ quantifies the departure from the Schwarzschild geometry. Considering $c_1$ as a perturbation, the leading order corrections to the metric function can be shown to be
\begin{equation}
    e^{\nu(r)} = \exp \left( \dfrac{c_1^2 r \left(a R^2 (4 r+R)+32 K_B\right)}{(4 r-R) (4 r+R)^2} \right) e^{\nu_\text{Schw}(r)}\,,
\end{equation}
and
\begin{equation}
    e^{\zeta(r)} = \exp \left( -\frac{c_1^2 \left(a R^2 (4 r+R)+16 K_B\right)}{4 (4 r+R)^2} \right) e^{\zeta_\text{Schw}(r)}\,,
\end{equation}
where $a$ is an integration constant, and $\nu_\text{Schw}(r)$ and $\zeta_\text{Schw}(r)$ are given by Eqs. \eqref{eq:schw_nu} and \eqref{eq:schw_zeta}.

We study the ${V}(r) = 0$ branch with ${F}(0,0)=0$ for asymptotic flatness. The metric equations become
\begin{equation}
\label{eq:tt_V0}
\begin{split}
    & -e^{\nu } \left(8 \zeta '+4 r \zeta ''+r \zeta^{\prime 2}\right) -2 e^{\zeta } \lambda  r \left(e^{\nu }+{U}^2\right)-2 K_B r {U}^{\prime 2} = 0 \,,
\end{split}
\end{equation}
and
\begin{equation}
\label{eq:rr_V0}
\begin{split}
    & e^{\nu } \left(4 \nu '+2 \zeta ' \left(r \nu '+2\right)+r \zeta^{\prime 2}\right) +2 e^{\zeta } \lambda  r \left(e^{\nu }-{U}^2\right)+2 K_B r {U}^{\prime 2} = 0 \,.
\end{split}
\end{equation}
The time component of the vector field equation becomes
\begin{equation}
\label{eq:vfe_V0}
    K_B \left(2 r {U}''+{U}' \left(r \zeta '-r \nu '+4\right)\right)-2 e^{\zeta } \lambda  r {U} = 0\,,
\end{equation}
while the unit-vector constraint reduces to
\begin{equation}
\label{eq:unit_vec_V0}
    e^{\nu }={U}^2 \,.
\end{equation}
The last two equations (Eqs. \eqref{eq:vfe_V0} and \eqref{eq:unit_vec_V0}) can be used to determine ${U}$ and the multiplier $\lambda$ in terms of the metric functions. Substituting these expressions into Eqs. \eqref{eq:tt_V0} and \eqref{eq:rr_V0} leads to
\begin{equation}
    8 \nu '+4 r \left(\zeta ''+\nu ''\right)+2 \zeta ' \left(r \nu '+4\right)+r \zeta ^{\prime 2}+r \nu^{\prime 2}=0\,,
\end{equation}
and
\begin{equation}
    \left(\zeta '+\nu '\right) \left(r \zeta '+r \nu '+4\right)=0 \,.
\end{equation}
There are two branches of solution here. The first one is
\begin{equation}
    e^{\zeta(r)} = \dfrac{e^{-\nu(r)}}{r^4}\,,
\end{equation}
while the second is
\begin{equation}
    e^{\zeta(r)} = e^{-\nu(r)}\,,
\end{equation}
where either $\nu$ or $\zeta$ can be chosen as a free function. The first one was presented in the main text (Eq. \eqref{eq:k2_metric}). Neither corresponds to a black hole solution.

\end{appendices}



\providecommand{\noopsort}[1]{}\providecommand{\singleletter}[1]{#1}%


\begin{thebibliography}{41}
\ifx \bisbn   \undefined \def \bisbn  #1{ISBN #1}\fi
\ifx \binits  \undefined \def \binits#1{#1}\fi
\ifx \bauthor  \undefined \def \bauthor#1{#1}\fi
\ifx \batitle  \undefined \def \batitle#1{#1}\fi
\ifx \bjtitle  \undefined \def \bjtitle#1{#1}\fi
\ifx \bvolume  \undefined \def \bvolume#1{\textbf{#1}}\fi
\ifx \byear  \undefined \def \byear#1{#1}\fi
\ifx \bissue  \undefined \def \bissue#1{#1}\fi
\ifx \bfpage  \undefined \def \bfpage#1{#1}\fi
\ifx \blpage  \undefined \def \blpage #1{#1}\fi
\ifx \burl  \undefined \def \burl#1{\textsf{#1}}\fi
\ifx \doiurl  \undefined \def \doiurl#1{\url{https://doi.org/#1}}\fi
\ifx \betal  \undefined \def \betal{\textit{et al.}}\fi
\ifx \binstitute  \undefined \def \binstitute#1{#1}\fi
\ifx \binstitutionaled  \undefined \def \binstitutionaled#1{#1}\fi
\ifx \bctitle  \undefined \def \bctitle#1{#1}\fi
\ifx \beditor  \undefined \def \beditor#1{#1}\fi
\ifx \bpublisher  \undefined \def \bpublisher#1{#1}\fi
\ifx \bbtitle  \undefined \def \bbtitle#1{#1}\fi
\ifx \bedition  \undefined \def \bedition#1{#1}\fi
\ifx \bseriesno  \undefined \def \bseriesno#1{#1}\fi
\ifx \blocation  \undefined \def \blocation#1{#1}\fi
\ifx \bsertitle  \undefined \def \bsertitle#1{#1}\fi
\ifx \bsnm \undefined \def \bsnm#1{#1}\fi
\ifx \bsuffix \undefined \def \bsuffix#1{#1}\fi
\ifx \bparticle \undefined \def \bparticle#1{#1}\fi
\ifx \barticle \undefined \def \barticle#1{#1}\fi
\bibcommenthead
\ifx \bconfdate \undefined \def \bconfdate #1{#1}\fi
\ifx \botherref \undefined \def \botherref #1{#1}\fi
\ifx \url \undefined \def \url#1{\textsf{#1}}\fi
\ifx \bchapter \undefined \def \bchapter#1{#1}\fi
\ifx \bbook \undefined \def \bbook#1{#1}\fi
\ifx \bcomment \undefined \def \bcomment#1{#1}\fi
\ifx \oauthor \undefined \def \oauthor#1{#1}\fi
\ifx \citeauthoryear \undefined \def \citeauthoryear#1{#1}\fi
\ifx \endbibitem  \undefined \def \endbibitem {}\fi
\ifx \bconflocation  \undefined \def \bconflocation#1{#1}\fi
\ifx \arxivurl  \undefined \def \arxivurl#1{\textsf{#1}}\fi
\csname PreBibitemsHook\endcsname

\bibitem{LIGOScientific:2016aoc}
\begin{barticle}
\bauthor{\bsnm{Abbott}, \binits{B.P.}}, \betal:
\batitle{{Observation of Gravitational Waves from a Binary Black Hole Merger}}.
\bjtitle{Phys. Rev. Lett.}
\bvolume{116}(\bissue{6}),
\bfpage{061102}
(\byear{2016})
{\href{https://arxiv.org/abs/1602.03837}{{arXiv:1602.03837}}}
{[gr-qc]}.
\doiurl{10.1103/PhysRevLett.116.061102}
\end{barticle}
\endbibitem

\bibitem{LIGOScientific:2020kqk}
\begin{barticle}
\bauthor{\bsnm{Abbott}, \binits{R.}}, \betal:
\batitle{{Population Properties of Compact Objects from the Second LIGO-Virgo
  Gravitational-Wave Transient Catalog}}.
\bjtitle{Astrophys. J. Lett.}
\bvolume{913}(\bissue{1}),
\bfpage{7}
(\byear{2021})
{\href{https://arxiv.org/abs/2010.14533}{{arXiv:2010.14533}}}
{[astro-ph.HE]}.
\doiurl{10.3847/2041-8213/abe949}
\end{barticle}
\endbibitem

\bibitem{EventHorizonTelescope:2019dse}
\begin{barticle}
\bauthor{\bsnm{Akiyama}, \binits{K.}}, \betal:
\batitle{{First M87 Event Horizon Telescope Results. I. The Shadow of the
  Supermassive Black Hole}}.
\bjtitle{Astrophys. J. Lett.}
\bvolume{875},
\bfpage{1}
(\byear{2019})
{\href{https://arxiv.org/abs/1906.11238}{{arXiv:1906.11238}}}
{[astro-ph.GA]}.
\doiurl{10.3847/2041-8213/ab0ec7}
\end{barticle}
\endbibitem

\bibitem{EventHorizonTelescope:2019ggy}
\begin{barticle}
\bauthor{\bsnm{Akiyama}, \binits{K.}}, \betal:
\batitle{{First M87 Event Horizon Telescope Results. VI. The Shadow and Mass of
  the Central Black Hole}}.
\bjtitle{Astrophys. J. Lett.}
\bvolume{875}(\bissue{1}),
\bfpage{6}
(\byear{2019})
{\href{https://arxiv.org/abs/1906.11243}{{arXiv:1906.11243}}}
{[astro-ph.GA]}.
\doiurl{10.3847/2041-8213/ab1141}
\end{barticle}
\endbibitem

\bibitem{Israel:1967wq}
\begin{barticle}
\bauthor{\bsnm{Israel}, \binits{W.}}:
\batitle{{Event horizons in static vacuum space-times}}.
\bjtitle{Phys. Rev.}
\bvolume{164},
\bfpage{1776}--\blpage{1779}
(\byear{1967}).
\doiurl{10.1103/PhysRev.164.1776}
\end{barticle}
\endbibitem

\bibitem{Israel:1967za}
\begin{barticle}
\bauthor{\bsnm{Israel}, \binits{W.}}:
\batitle{{Event horizons in static electrovac space-times}}.
\bjtitle{Commun. Math. Phys.}
\bvolume{8},
\bfpage{245}--\blpage{260}
(\byear{1968}).
\doiurl{10.1007/BF01645859}
\end{barticle}
\endbibitem

\bibitem{Carter:1971zc}
\begin{barticle}
\bauthor{\bsnm{Carter}, \binits{B.}}:
\batitle{{Axisymmetric Black Hole Has Only Two Degrees of Freedom}}.
\bjtitle{Phys. Rev. Lett.}
\bvolume{26},
\bfpage{331}--\blpage{333}
(\byear{1971}).
\doiurl{10.1103/PhysRevLett.26.331}
\end{barticle}
\endbibitem

\bibitem{Bekenstein:1995un}
\begin{barticle}
\bauthor{\bsnm{Bekenstein}, \binits{J.D.}}:
\batitle{{Novel
  \textquoteleft{}\textquoteleft{}no-scalar-hair\textquoteright{}\textquoteright{}
  theorem for black holes}}.
\bjtitle{Phys. Rev. D}
\bvolume{51}(\bissue{12}),
\bfpage{6608}
(\byear{1995}).
\doiurl{10.1103/PhysRevD.51.R6608}
\end{barticle}
\endbibitem

\bibitem{Hertog:2006rr}
\begin{barticle}
\bauthor{\bsnm{Hertog}, \binits{T.}}:
\batitle{{Towards a Novel no-hair Theorem for Black Holes}}.
\bjtitle{Phys. Rev. D}
\bvolume{74},
\bfpage{084008}
(\byear{2006})
{\href{https://arxiv.org/abs/gr-qc/0608075}{{arXiv:gr-qc/0608075}}}.
\doiurl{10.1103/PhysRevD.74.084008}
\end{barticle}
\endbibitem

\bibitem{Hui:2012qt}
\begin{barticle}
\bauthor{\bsnm{Hui}, \binits{L.}},
\bauthor{\bsnm{Nicolis}, \binits{A.}}:
\batitle{{No-Hair Theorem for the Galileon}}.
\bjtitle{Phys. Rev. Lett.}
\bvolume{110},
\bfpage{241104}
(\byear{2013})
{\href{https://arxiv.org/abs/1202.1296}{{arXiv:1202.1296}}}
{[hep-th]}.
\doiurl{10.1103/PhysRevLett.110.241104}
\end{barticle}
\endbibitem

\bibitem{Sotiriou:2015pka}
\begin{barticle}
\bauthor{\bsnm{Sotiriou}, \binits{T.P.}}:
\batitle{{Black Holes and Scalar Fields}}.
\bjtitle{Class. Quant. Grav.}
\bvolume{32}(\bissue{21}),
\bfpage{214002}
(\byear{2015})
{\href{https://arxiv.org/abs/1505.00248}{{arXiv:1505.00248}}}
{[gr-qc]}.
\doiurl{10.1088/0264-9381/32/21/214002}
\end{barticle}
\endbibitem

\bibitem{Barack:2018yly}
\begin{barticle}
\bauthor{\bsnm{Barack}, \binits{L.}}, \betal:
\batitle{{Black holes, gravitational waves and fundamental physics: a
  roadmap}}.
\bjtitle{Class. Quant. Grav.}
\bvolume{36}(\bissue{14}),
\bfpage{143001}
(\byear{2019})
{\href{https://arxiv.org/abs/1806.05195}{{arXiv:1806.05195}}}
{[gr-qc]}.
\doiurl{10.1088/1361-6382/ab0587}
\end{barticle}
\endbibitem

\bibitem{Barausse:2020rsu}
\begin{barticle}
\bauthor{\bsnm{Barausse}, \binits{E.}}, \betal:
\batitle{{Prospects for Fundamental Physics with LISA}}.
\bjtitle{Gen. Rel. Grav.}
\bvolume{52}(\bissue{8}),
\bfpage{81}
(\byear{2020})
{\href{https://arxiv.org/abs/2001.09793}{{arXiv:2001.09793}}}
{[gr-qc]}.
\doiurl{10.1007/s10714-020-02691-1}
\end{barticle}
\endbibitem

\bibitem{Milgrom:1983ca}
\begin{barticle}
\bauthor{\bsnm{Milgrom}, \binits{M.}}:
\batitle{{A Modification of the Newtonian dynamics as a possible alternative to
  the hidden mass hypothesis}}.
\bjtitle{Astrophys. J.}
\bvolume{270},
\bfpage{365}--\blpage{370}
(\byear{1983}).
\doiurl{10.1086/161130}
\end{barticle}
\endbibitem

\bibitem{Milgrom:1983pn}
\begin{barticle}
\bauthor{\bsnm{Milgrom}, \binits{M.}}:
\batitle{{A Modification of the Newtonian dynamics: Implications for
  galaxies}}.
\bjtitle{Astrophys. J.}
\bvolume{270},
\bfpage{371}--\blpage{383}
(\byear{1983}).
\doiurl{10.1086/161131}
\end{barticle}
\endbibitem

\bibitem{Milgrom:1983zz}
\begin{barticle}
\bauthor{\bsnm{Milgrom}, \binits{M.}}:
\batitle{{A modification of the Newtonian dynamics: implications for galaxy
  systems}}.
\bjtitle{Astrophys. J.}
\bvolume{270},
\bfpage{384}--\blpage{389}
(\byear{1983}).
\doiurl{10.1086/161132}
\end{barticle}
\endbibitem

\bibitem{Bekenstein:2004ne}
\begin{barticle}
\bauthor{\bsnm{Bekenstein}, \binits{J.D.}}:
\batitle{{Relativistic gravitation theory for the MOND paradigm}}.
\bjtitle{Phys. Rev. D}
\bvolume{70},
\bfpage{083509}
(\byear{2004})
{\href{https://arxiv.org/abs/astro-ph/0403694}{{arXiv:astro-ph/0403694}}}.
\doiurl{10.1103/PhysRevD.70.083509}.
\bcomment{[Erratum: Phys.Rev.D 71, 069901 (2005)]}
\end{barticle}
\endbibitem

\bibitem{Sanders:2005vd}
\begin{barticle}
\bauthor{\bsnm{Sanders}, \binits{R.H.}}:
\batitle{{A Tensor-vector-scalar framework for modified dynamics and cosmic
  dark matter}}.
\bjtitle{Mon. Not. Roy. Astron. Soc.}
\bvolume{363},
\bfpage{459}
(\byear{2005})
{\href{https://arxiv.org/abs/astro-ph/0502222}{{arXiv:astro-ph/0502222}}}.
\doiurl{10.1111/j.1365-2966.2005.09375.x}
\end{barticle}
\endbibitem

\bibitem{Skordis:2005xk}
\begin{barticle}
\bauthor{\bsnm{Skordis}, \binits{C.}},
\bauthor{\bsnm{Mota}, \binits{D.F.}},
\bauthor{\bsnm{Ferreira}, \binits{P.G.}},
\bauthor{\bsnm{Boehm}, \binits{C.}}:
\batitle{{Large Scale Structure in Bekenstein's theory of relativistic Modified
  Newtonian Dynamics}}.
\bjtitle{Phys. Rev. Lett.}
\bvolume{96},
\bfpage{011301}
(\byear{2006})
{\href{https://arxiv.org/abs/astro-ph/0505519}{{arXiv:astro-ph/0505519}}}.
\doiurl{10.1103/PhysRevLett.96.011301}
\end{barticle}
\endbibitem

\bibitem{Zlosnik:2006sb}
\begin{barticle}
\bauthor{\bsnm{Zlosnik}, \binits{T.G.}},
\bauthor{\bsnm{Ferreira}, \binits{P.G.}},
\bauthor{\bsnm{Starkman}, \binits{G.D.}}:
\batitle{{The Vector-tensor nature of Bekenstein's relativistic theory of
  modified gravity}}.
\bjtitle{Phys. Rev. D}
\bvolume{74},
\bfpage{044037}
(\byear{2006})
{\href{https://arxiv.org/abs/gr-qc/0606039}{{arXiv:gr-qc/0606039}}}.
\doiurl{10.1103/PhysRevD.74.044037}
\end{barticle}
\endbibitem

\bibitem{Babichev:2011kq}
\begin{barticle}
\bauthor{\bsnm{Babichev}, \binits{E.}},
\bauthor{\bsnm{Deffayet}, \binits{C.}},
\bauthor{\bsnm{Esposito-Farese}, \binits{G.}}:
\batitle{{Improving relativistic MOND with Galileon k-mouflage}}.
\bjtitle{Phys. Rev. D}
\bvolume{84},
\bfpage{061502}
(\byear{2011})
{\href{https://arxiv.org/abs/1106.2538}{{arXiv:1106.2538}}}
{[gr-qc]}.
\doiurl{10.1103/PhysRevD.84.061502}
\end{barticle}
\endbibitem

\bibitem{Famaey:2011kh}
\begin{barticle}
\bauthor{\bsnm{Famaey}, \binits{B.}},
\bauthor{\bsnm{McGaugh}, \binits{S.}}:
\batitle{{Modified Newtonian Dynamics (MOND): Observational Phenomenology and
  Relativistic Extensions}}.
\bjtitle{Living Rev. Rel.}
\bvolume{15},
\bfpage{10}
(\byear{2012})
{\href{https://arxiv.org/abs/1112.3960}{{arXiv:1112.3960}}}
{[astro-ph.CO]}.
\doiurl{10.12942/lrr-2012-10}
\end{barticle}
\endbibitem

\bibitem{Zlosnik:2017xpr}
\begin{barticle}
\bauthor{\bsnm{Z\l{}o\'snik}, \binits{T.G.}},
\bauthor{\bsnm{Skordis}, \binits{C.}}:
\batitle{{Cosmology of the Galileon extension of Bekenstein\textquoteright{}s
  theory of relativistic modified Newtonian dynamics}}.
\bjtitle{Phys. Rev. D}
\bvolume{95}(\bissue{12}),
\bfpage{124023}
(\byear{2017})
{\href{https://arxiv.org/abs/1702.00683}{{arXiv:1702.00683}}}
{[gr-qc]}.
\doiurl{10.1103/PhysRevD.95.124023}
\end{barticle}
\endbibitem

\bibitem{PhysRevD.100.104013}
\begin{barticle}
\bauthor{\bsnm{Skordis}, \binits{C.}},
\bauthor{\bsnm{Zlosnik}, \binits{T.}}:
\batitle{Gravitational alternatives to dark matter with tensor mode speed
  equaling the speed of light}.
\bjtitle{Phys. Rev. D}
\bvolume{100},
\bfpage{104013}
(\byear{2019}).
\doiurl{10.1103/PhysRevD.100.104013}
\end{barticle}
\endbibitem

\bibitem{Skordis:2020eui}
\begin{barticle}
\bauthor{\bsnm{Skordis}, \binits{C.}},
\bauthor{\bsnm{Zlosnik}, \binits{T.}}:
\batitle{{New Relativistic Theory for Modified Newtonian Dynamics}}.
\bjtitle{Phys. Rev. Lett.}
\bvolume{127}(\bissue{16}),
\bfpage{161302}
(\byear{2021})
{\href{https://arxiv.org/abs/2007.00082}{{arXiv:2007.00082}}}
{[astro-ph.CO]}.
\doiurl{10.1103/PhysRevLett.127.161302}
\end{barticle}
\endbibitem

\bibitem{Skordis:2021vuk}
\begin{botherref}
\oauthor{\bsnm{Skordis}, \binits{C.}},
\oauthor{\bsnm{Zlosnik}, \binits{T.}}:
{Linear stability of the new relativistic theory of modified Newtonian
  dynamics}
(2021)
{\href{https://arxiv.org/abs/2109.13287}{{arXiv:2109.13287}}}
{[gr-qc]}
\end{botherref}
\endbibitem

\bibitem{LIGOScientific:2021sio}
\begin{botherref}
\oauthor{\bsnm{Abbott}, \binits{R.}}, et al.:
{Tests of General Relativity with GWTC-3}
(2021)
{\href{https://arxiv.org/abs/2112.06861}{{arXiv:2112.06861}}}
{[gr-qc]}
\end{botherref}
\endbibitem

\bibitem{reggie_bernardo_4810864}
\begin{botherref}
\oauthor{\bsnm{Bernardo}, \binits{R.}}:
\href{https://doi.org/10.5281/zenodo.4810864}{reggiebernardo/notebooks: dark
  energy research notebooks}.
\href{https://zenodo.org/}{Zenodo}
(2021)
\end{botherref}
\endbibitem

\bibitem{Kashfi:2022dyb}
\begin{botherref}
\oauthor{\bsnm{Kashfi}, \binits{T.}},
\oauthor{\bsnm{Roshan}, \binits{M.}}:
{Cosmological Dynamics of Relativistic MOND}
(2022)
{\href{https://arxiv.org/abs/2204.05672}{{arXiv:2204.05672}}}
{[gr-qc]}
\end{botherref}
\endbibitem

\bibitem{Giannios:2005es}
\begin{barticle}
\bauthor{\bsnm{Giannios}, \binits{D.}}:
\batitle{{Spherically symmetric, static spacetimes in TeVeS}}.
\bjtitle{Phys. Rev. D}
\bvolume{71},
\bfpage{103511}
(\byear{2005})
{\href{https://arxiv.org/abs/gr-qc/0502122}{{arXiv:gr-qc/0502122}}}.
\doiurl{10.1103/PhysRevD.71.103511}
\end{barticle}
\endbibitem

\bibitem{Bernardo:2019mmx}
\begin{barticle}
\bauthor{\bsnm{Bernardo}, \binits{R.C.}},
\bauthor{\bsnm{Vega}, \binits{I.}}:
\batitle{{Hair-dressing Horndeski: An approach for obtaining hairy solutions in
  cubic Horndeski gravity}}.
\bjtitle{Phys. Rev. D}
\bvolume{99}(\bissue{12}),
\bfpage{124049}
(\byear{2019})
{\href{https://arxiv.org/abs/1902.04988}{{arXiv:1902.04988}}}
{[gr-qc]}.
\doiurl{10.1103/PhysRevD.99.124049}
\end{barticle}
\endbibitem

\bibitem{Bahamonde:2021gfp}
\begin{botherref}
\oauthor{\bsnm{Bahamonde}, \binits{S.}},
\oauthor{\bsnm{Dialektopoulos}, \binits{K.F.}},
\oauthor{\bsnm{Escamilla-Rivera}, \binits{C.}},
\oauthor{\bsnm{Farrugia}, \binits{G.}},
\oauthor{\bsnm{Gakis}, \binits{V.}},
\oauthor{\bsnm{Hendry}, \binits{M.}},
\oauthor{\bsnm{Hohmann}, \binits{M.}},
\oauthor{\bsnm{Said}, \binits{J.L.}},
\oauthor{\bsnm{Mifsud}, \binits{J.}},
\oauthor{\bsnm{Di~Valentino}, \binits{E.}}:
{Teleparallel Gravity: From Theory to Cosmology}
(2021)
{\href{https://arxiv.org/abs/2106.13793}{{arXiv:2106.13793}}}
{[gr-qc]}
\end{botherref}
\endbibitem

\bibitem{Babichev:2016fbg}
\begin{barticle}
\bauthor{\bsnm{Babichev}, \binits{E.}},
\bauthor{\bsnm{Charmousis}, \binits{C.}},
\bauthor{\bsnm{Leh\'ebel}, \binits{A.}},
\bauthor{\bsnm{Moskalets}, \binits{T.}}:
\batitle{{Black holes in a cubic Galileon universe}}.
\bjtitle{JCAP}
\bvolume{09},
\bfpage{011}
(\byear{2016})
{\href{https://arxiv.org/abs/1605.07438}{{arXiv:1605.07438}}}
{[gr-qc]}.
\doiurl{10.1088/1475-7516/2016/09/011}
\end{barticle}
\endbibitem

\bibitem{Babichev:2017guv}
\begin{barticle}
\bauthor{\bsnm{Babichev}, \binits{E.}},
\bauthor{\bsnm{Charmousis}, \binits{C.}},
\bauthor{\bsnm{Leh\'ebel}, \binits{A.}}:
\batitle{{Asymptotically flat black holes in Horndeski theory and beyond}}.
\bjtitle{JCAP}
\bvolume{04},
\bfpage{027}
(\byear{2017})
{\href{https://arxiv.org/abs/1702.01938}{{arXiv:1702.01938}}}
{[gr-qc]}.
\doiurl{10.1088/1475-7516/2017/04/027}
\end{barticle}
\endbibitem

\bibitem{Babichev:2016rlq}
\begin{barticle}
\bauthor{\bsnm{Babichev}, \binits{E.}},
\bauthor{\bsnm{Charmousis}, \binits{C.}},
\bauthor{\bsnm{Leh\'ebel}, \binits{A.}}:
\batitle{{Black holes and stars in Horndeski theory}}.
\bjtitle{Class. Quant. Grav.}
\bvolume{33}(\bissue{15}),
\bfpage{154002}
(\byear{2016})
{\href{https://arxiv.org/abs/1604.06402}{{arXiv:1604.06402}}}
{[gr-qc]}.
\doiurl{10.1088/0264-9381/33/15/154002}
\end{barticle}
\endbibitem

\bibitem{R:2022cwe}
\begin{barticle}
\bauthor{\bsnm{R.}, \binits{A.H.K.}},
\bauthor{\bsnm{Most}, \binits{E.R.}},
\bauthor{\bsnm{Noronha}, \binits{J.}},
\bauthor{\bsnm{Witek}, \binits{H.}},
\bauthor{\bsnm{Yunes}, \binits{N.}}:
\batitle{{How do spherical black holes grow monopole hair?}}
\bjtitle{Phys. Rev. D}
\bvolume{105}(\bissue{6}),
\bfpage{064041}
(\byear{2022})
{\href{https://arxiv.org/abs/2201.05178}{{arXiv:2201.05178}}}
{[gr-qc]}.
\doiurl{10.1103/PhysRevD.105.064041}
\end{barticle}
\endbibitem

\bibitem{rcb_cyc_in_prep}
\begin{botherref}
\oauthor{\bsnm{Bernardo}, \binits{R.C.}},
\oauthor{\bsnm{Chen}, \binits{C.-Y.}}:
In preparation
(2022)
\end{botherref}
\endbibitem

\bibitem{Zhang:2021ygh}
\begin{barticle}
\bauthor{\bsnm{Zhang}, \binits{Z.}}:
\batitle{{Geometrization of light bending and its application to SdS$_{w}$
  spacetime}}.
\bjtitle{Class. Quant. Grav.}
\bvolume{39}(\bissue{1}),
\bfpage{015003}
(\byear{2022})
{\href{https://arxiv.org/abs/2112.04149}{{arXiv:2112.04149}}}
{[gr-qc]}.
\doiurl{10.1088/1361-6382/ac38d1}
\end{barticle}
\endbibitem

\bibitem{Tattersall:2018nve}
\begin{barticle}
\bauthor{\bsnm{Tattersall}, \binits{O.J.}},
\bauthor{\bsnm{Ferreira}, \binits{P.G.}}:
\batitle{{Quasinormal modes of black holes in Horndeski gravity}}.
\bjtitle{Phys. Rev. D}
\bvolume{97}(\bissue{10}),
\bfpage{104047}
(\byear{2018})
{\href{https://arxiv.org/abs/1804.08950}{{arXiv:1804.08950}}}
{[gr-qc]}.
\doiurl{10.1103/PhysRevD.97.104047}
\end{barticle}
\endbibitem

\bibitem{Lagos:2020oek}
\begin{barticle}
\bauthor{\bsnm{Lagos}, \binits{M.}},
\bauthor{\bsnm{Ferreira}, \binits{P.G.}},
\bauthor{\bsnm{Tattersall}, \binits{O.J.}}:
\batitle{{Anomalous decay rate of quasinormal modes}}.
\bjtitle{Phys. Rev. D}
\bvolume{101}(\bissue{8}),
\bfpage{084018}
(\byear{2020})
{\href{https://arxiv.org/abs/2002.01897}{{arXiv:2002.01897}}}
{[gr-qc]}.
\doiurl{10.1103/PhysRevD.101.084018}
\end{barticle}
\endbibitem

\bibitem{Bernardo:2021vsj}
\begin{barticle}
\bauthor{\bsnm{Bernardo}, \binits{R.C.}}:
\batitle{{Gravitational wave signatures from dark sector interactions}}.
\bjtitle{Phys. Rev. D}
\bvolume{104}(\bissue{2}),
\bfpage{024070}
(\byear{2021})
{\href{https://arxiv.org/abs/2103.02311}{{arXiv:2103.02311}}}
{[gr-qc]}.
\doiurl{10.1103/PhysRevD.104.024070}
\end{barticle}
\endbibitem

\end{thebibliography}


\end{document}